\documentclass[11pt]{article}

\DeclareMathAlphabet{\scr}{U}{rsfs}{m}{n}
\usepackage{latexsym}
\usepackage{epsfig}
\usepackage[mathscr]{eucal}
\usepackage{amsfonts}
\usepackage{amscd}
\usepackage{cite}
\usepackage{array}   
\usepackage{amssymb}
\usepackage{colordvi}
\usepackage[centertags]{amsmath}
\usepackage{enumerate}
\usepackage{graphicx}
\usepackage{booktabs}
\usepackage{theorem}
\usepackage[footnotesize]{caption}
\usepackage{soul}
\usepackage{mcite}
\usepackage{slashed}
\usepackage{color}
\usepackage{ulem}
\usepackage{bbm}
\usepackage[utf8]{inputenc}
\newcommand{\newc}{\newcommand}
\newc{\be}{\begin{equation}}
\newc{\ee}{\end{equation}}
\newc{\bea}{\begin{eqnarray}}
\newc{\eea}{\end{eqnarray}}
\newc{\ol}{\overline}
\newc{\wt}{\widetilde}
\newc{\bs}{\boldsymbol}
\newc{\m}{\mathcal}
\newc{\la}{\langle}
\newc{\ra}{\rangle}

\newcommand{\beq}{\begin{eqnarray}} 
\newcommand{\eeq}{\end{eqnarray}} 
\newcommand{\bpmatrix}{\begin{pmatrix}}
\newcommand{\epmatrix}{\end{pmatrix}}
\newcommand{\ba}{\begin{array}}
\newcommand{\ea}{\end{array}}

\renewcommand{\ol}{\text{1l}}

\renewcommand{\eqref}[1]{Eq.~(\ref{#1})}

\newcommand{\bc}{\begin{center}}
\newcommand{\ec}{\end{center}}

\newcommand{\gsim}{\raisebox{-0.13cm}{~\shortstack{$>$ \\[-0.07cm]
      $\sim$}}~}
\newcommand{\lsim}{\raisebox{-0.13cm}{~\shortstack{$<$ \\[-0.07cm]
      $\sim$}}~}
\newcommand{\s}{\newline \vspace*{-3.5mm}}

\begin{document}

\title{
\vspace*{-3cm}
\phantom{h} \hfill\mbox{\small KA-TP-31-2016}
\\[1cm]
\textbf{2HDM Higgs-to-Higgs Decays \\
 at Next-to-Leading Order \\[4mm]}}

\date{}
\author{
Marcel Krause$^{1\,}$\footnote{E-mail:
  \texttt{marcel.krause@mail.de}} ,
Margarete M\"{u}hlleitner$^{1\,}$\footnote{E-mail:
  \texttt{margarete.muehlleitner@kit.edu}} ,
Rui Santos$^{2\,, \, 3\,}$\footnote{E-mail:
  \texttt{rasantos@fc.ul.pt}} ,
Hanna Ziesche$^{1\,}$\footnote{E-mail: \texttt{hanna.ziesche@kit.edu}}
\\[9mm]
{\small\it
$^1$Institute for Theoretical Physics, Karlsruhe Institute of Technology,} \\
{\small\it 76128 Karlsruhe, Germany}\\[3mm]
{\small\it
$^2$ISEL -
 Instituto Superior de Engenharia de Lisboa,} \\
{\small \it  Instituto Polit\'ecnico de Lisboa,  1959-007 Lisboa, Portugal}\\[3mm]
{\small\it
$^3$Centro de F\'{\i}sica Te\'{o}rica e Computacional,
    Faculdade de Ci\^{e}ncias,} \\
{\small \it    Universidade de Lisboa, Campo Grande, Edif\'{\i}cio C8
  1749-016 Lisboa, Portugal} 
}

\maketitle

\begin{abstract}
\noindent
The detailed investigation of the Higgs sector at present and future
colliders necessitates from the theory side as precise predictions as possible,
including higher order corrections. An important ingredient for the
computation of higher order corrections is the renormalization of the
model parameters and fields. In this paper we complete  
the renormalization of the 2-Higgs-Doublet Model (2HDM) Higgs sector
launched in a previous contribution with the investigation of the
renormalization of the mixing angles $\alpha$ and $\beta$. Here, we treat the
renormalization of the mass parameter $m_{12}^2$ that softly breaks the
$\mathbb{Z}_2$ symmetry of the 2HDM Higgs sector. We investigate the
impact of two different renormalization schemes on the sample
Higgs-to-Higgs decay $H\to hh$. This decay also allows us to
analyze the renormalization of the mixing angles and to confirm the 
properties extracted before in other Higgs decays. In
conclusion we find that a gauge-independent, process-independent
and numerically stable renormalization of the 2HDM Higgs sector
is given by the application of the tadpole-pinched scheme for the
mixing angles $\alpha$ and $\beta$ and by the use of the
$\overline{\mbox{MS}}$ scheme for $m_{12}^2$.
\end{abstract}
\thispagestyle{empty}
\vfill
\newpage
\setcounter{page}{1}

\section{Introduction}
The experimental data \cite{Khachatryan:2014kca,Aad:2015mxa,Khachatryan:2014jba,Aad:2015gba} on the properties of the Higgs boson discovered in 2012 by the LHC
experiments ATLAS \cite{Aad:2012tfa} and CMS
\cite{Chatrchyan:2012ufa} are compatible with a Standard Model
(SM)-like Higgs boson. Still they leave room for interpretations in
models beyond the SM 
(BSM). Theoretical and experimental considerations lead to the
conclusion that the SM cannot be the ultimate theory of nature. In
view of no direct discovery of BSM manifestations in form of new
particles so far, we are
bound to study the Higgs sector in great detail in order to gain
insights in possibly existing new physics (NP). Among the plethora of
BSM extensions of the Higgs sector, 2-Higgs-Doublet Models (2HDM)
\cite{Gunion:1989we,Lee:1973iz,Branco:2011iw} play an important
role. They feature five physical Higgs bosons, two CP-even ones
$h$ and $H$, a CP-odd scalar $A$ and two charged Higgs bosons
$H^\pm$. The couplings of these Higgs bosons to SM particles are
modified by two mixing angles, the angle $\alpha$ arising from the
diagonalization of the CP-even Higgs mass matrix, and $\beta$
originating from the CP-odd and charged Higgs sectors. 
Together with singlet models, 2HDMs form the simplest SM extensions
that are compatible with theoretical and experimental constraints
\cite{Barroso:2013zxa,Ferreira:2013qua,Dumont:2014wha,Bernon:2014vta,Dumont:2014kna}. Additionally,
the Higgs sector of the Minimal Supersymmetric Extension of the SM (MSSM)
\cite{Gunion:1989we,Martin:1997ns,Dawson:1997tz,Djouadi:2005gj}
represents a special case of the 2HDM type II. This allows to map
insights gained in investigations of the 
2HDM onto the MSSM and to compare effects that are possible in the less restricted
2HDM to the situation in the more restrained supersymmetric Higgs
sector. The comparison of different models and, ultimately, the identification of
the underlying theory requires experimental data at highest precision.
Besides excellent experimental analysis techniques and the accumulation of
a large amount of data at sufficiently high energy, this necessitates
from the theory side precise predictions on observables and
parameters, including higher order corrections. In a previous paper
\cite{Krause:2016oke} we have provided an important basis for the
computation of higher order (HO) corrections in the 2HDM by working
out a manifestly gauge-independent renormalization of the two 2HDM
mixing angles $\alpha$ and $\beta$, which is additionally
process independent and numerically
stable.\footnote{Recently, in \cite{Denner:2016etu} an
    $\overline{\mbox{MS}}$ scheme was proposed for $\alpha$ and
    $\beta$.} The mixing angles play an 
important role for phenomenology, and we have investigated our
renormalization scheme in the sample decays of the charged Higgs boson
into a $W$ boson and a CP-even scalar, $H^\pm \to W^\pm h/H$, and of
the heavy CP-even Higgs decay into a pair of $Z$ bosons, $H\to ZZ$. \s 

In this paper we complete our renormalization of the 2HDM Higgs sector
by computing the next-to-leading order (NLO) corrections to
Higgs-to-Higgs decays. 
The investigation of these decays is of particular phenomenological
interest. Not only they are a clear manifestation of an extended Higgs
sector, they also give access to the trilinear Higgs
self-couplings. The determination of these self-interactions
constitutes a first important step towards the reconstruction of the
Higgs potential
\cite{Djouadi:1999gv,Djouadi:1999rca,Muhlleitner:2000jj}, which is the
final missing piece in the experimental verification of the Higgs
mechanism. Higgs-to-Higgs 
decays can also be exploited for the discovery of non-SM Higgs bosons through
cascade decays that are not accessible directly (see
e.g.~\cite{Nhung:2013lpa,No:2013wsa,Arhrib:2013oia,Baglio:2014nea,King:2014xwa,Barger:2014qva,Bomark:2014gya}). Interestingly, they can also be used to
distinguish between different models \cite{Costa:2015llh}. It might
even be that we see NP in Higgs pair production before anywhere else,
{\it i.e.}~in particular for Higgs couplings of the 125 GeV resonance which are
SM-like \cite{Grober:2016wmf}. \s

Compared to the NLO computation of the Higgs decays presented
in \cite{Krause:2016oke}, the HO corrections to Higgs-to-Higgs decays
require in addition the renormalization
of the mass parameter $m_{12}^2$ of the Higgs potential. This
parameter softly breaks the discrete $\mathbb{Z}_2$ symmetry, imposed
to avoid tree-level flavour changing neutral currents (FCNC). 
We suggest different renormalization schemes for $m_{12}^2$ and
investigate their numerical stability with respect to typical sizes of higher order
corrections encountered in 2HDM Higgs-to-Higgs decays. The sample
decay chosen in our analysis additionally allows us to study the
numerical stability of the angular renormalization schemes proposed in
\cite{Krause:2016oke} in a process which shows in the Higgs self-coupling a
much more involved dependence on the mixing angles than the previously
studied decays. In order to do so we identify the 2HDM
parameter regions that lead to parametrically enhanced loop
corrections due to non-decoupling effects. Subsequently, we analyze
the loop corrections with respect to numerical stability in the
decoupling regime where the heavy Higgs masses are due to a large mass
scale independently of the Higgs self-couplings. \s

The paper is organized as follows. In section~\ref{sec:modeldesc} we
briefly introduce our model and set the notation. In the following
section~\ref{sec:renormcond} we shortly review our renormalization conditions
of \cite{Krause:2016oke}, also needed here, and introduce the additionally required
renormalization of $m_{12}^2$ entering the
loop-corrected Higgs-to-Higgs decays. In section~\ref{sec:nlocalc} we
describe the calculation of the electroweak one-loop correction to the sample
decay $H \to hh$. The numerical analysis is presented in
section~\ref{sec:numerics} in which we investigate our proposed
renormalization procedures with respect to gauge independence, process
independence and numerical stability. Our conclusions are given in
section~\ref{sec:concl}. 

\section{Description of the Model \label{sec:modeldesc}}
\setcounter{equation}{0}
Our work is performed within the framework of a general 2HDM with a
global softly broken discrete $\mathbb{Z}_2$ symmetry. For the kinetic
term of the two $SU(2)_L$ Higgs doublets $\Phi_1$ and $\Phi_2$ we
introduce the covariant derivative
\beq
D_\mu = \partial_\mu + \frac{i}{2} g \sum_{a=1}^3 \tau^a W_\mu^a +
\frac{i}{2} g' B_\mu \;, \label{eq:covdiv}
\eeq
where $\tau^a$ denote the Pauli matrices, $W_\mu^a$ and $B_\mu$ the
$SU(2)_L$ and $U(1)_Y$ gauge bosons, respectively, and $g$ and $g'$ the
corresponding gauge couplings. The Higgs sector is
described by the kinetic Lagrangian 
\beq
{\cal L}_{\text{kin}} = \sum_{i=1}^2 (D_\mu \Phi_i)^\dagger (D^\mu \Phi_i)
\eeq
and the scalar potential, which can be cast into the form 
\beq
V &=& m_{11}^2 |\Phi_1|^2 + m_{22}^2 |\Phi_2|^2 - m_{12}^2 (\Phi_1^\dagger
\Phi_2 + h.c.) + \frac{\lambda_1}{2} (\Phi_1^\dagger \Phi_1)^2 +
\frac{\lambda_2}{2} (\Phi_2^\dagger \Phi_2)^2 \nonumber \\
&& + \lambda_3
(\Phi_1^\dagger \Phi_1) (\Phi_2^\dagger \Phi_2) + \lambda_4 
(\Phi_1^\dagger \Phi_2) (\Phi_2^\dagger \Phi_1) + \frac{\lambda_5}{2}
[(\Phi_1^\dagger \Phi_2)^2 + h.c.] \;.
\eeq
The absence of FCNCs at tree 
level is ensured by imposing the discrete $\mathbb{Z}_2$ symmetry
under which the doublets transform as $\Phi_1 \to -\Phi_1$ and $\Phi_2 \to
\Phi_2$. We assume CP conservation so that the 2HDM potential depends
on eight real parameters, three mass parameters, $m_{11}$, $m_{22}$
and $m_{12}$, and five coupling parameters
$\lambda_1$-$\lambda_5$. As can be inferred from the potential, a
non-zero value of $m_{12}^2$ softly breaks the discrete $\mathbb{Z}_2$
symmetry. 
The two doublet fields $\Phi_1$ and $\Phi_2$ can be expressed in terms
of charged complex fields $\phi_i^+$ and real neutral CP-even and
CP-odd fields $\rho_i$ and $\eta_i$ ($i=1,2$), respectively. 
By expanding the two Higgs doublets about their vacuum expectation
values (VEVs), developed after electroweak symmetry breaking (EWSB),
which are real in the CP-conserving case, 
\beq
\Phi_1 = \left(
\begin{array}{c}
\phi_1^+ \\
\frac{\rho_1 + i \eta_1 + v_1}{\sqrt{2}}
\end{array}
\right) \qquad \mbox{and} \qquad
\Phi_2 = \left(
\begin{array}{c}
\phi_2^+ \\
\frac{\rho_2 + i \eta_2 + v_2}{\sqrt{2}}
\end{array}
\right) \;, \label{eq:vevexpansion}
\eeq
the mass matrices can be derived from the terms bilinear in the Higgs
fields in the Higgs potential. Under the assumption of charge and CP
conservation they decompose into $2 \times 2$ matrices ${\cal M}_S$,
${\cal M}_P$ and ${\cal M}_C$ for the neutral CP-even, neutral CP-odd
and charged Higgs sectors.
For the two Higgs doublets $\Phi_i$ to take their minimum at $\langle
\Phi_i \rangle \equiv v_i$ the minimum conditions 
\beq
\left.\frac{\partial V}{\partial \Phi_1}\right|_{\langle \Phi_i \rangle} = 
\left.\frac{\partial V}{\partial \Phi_2}\right|_{\langle \Phi_i
  \rangle} = 0 \;, \label{eq:tadcond} 
\eeq
have to be fulfilled. This is equivalent to the requirement of the two terms
linear in $\rho_1$ and $\rho_2$ to vanish, {\it i.e.}
\beq
\frac{T_1}{v_1} &=&  m_{11}^2 - m_{12}^2
\frac{v_2}{v_1} + \frac{\lambda_1 v_1^2}{2} 
+ \frac{\lambda_{345} v_2^2}{2} \label{eq:tad1} \\
\frac{T_2}{v_2} &=&  m_{22}^2 - m_{12}^2
\frac{v_1}{v_2} + \frac{\lambda_2 v_2^2}{2} + \frac{\lambda_{345}
  v_1^2}{2} \label{eq:tad2} \;.
\eeq
The tadpole conditions can be exploited to replace $m_{11}^2$ and
$m_{22}^2$ by the tadpole parameters $T_1$ and $T_2$. This yields the
following explicit form of the mass matrices 
\beq
{\cal M}_S &=& \left( \begin{array}{cc} m_{12}^2 \frac{v_2}{v_1} +
    \lambda_1 v_1^2 & -
    m_{12}^2 + \lambda_{345} v_1 v_2 \\ -m_{12}^2 + \lambda_{345} v_1
    v_2 & m_{12}^2 \frac{v_1}{v_2} + \lambda_2
    v_2^2 \end{array}\right) + \left( \begin{array}{cc}
    \frac{T_1}{v_1} & 0 \\ 0 & \frac{T_2}{v_2} \end{array}\right)  
\label{eq:scalarmass} \\ 
{\cal M}_P &=& \left( \frac{m_{12}^2}{v_1 v_2} - \lambda_5 \right)
\left( \begin{array}{cc} v_2^2 & - v_1 v_2 \\ - v_1 v_2 &
    v_1^2 \end{array} \right) + \left( \begin{array}{cc}
    \frac{T_1}{v_1} & 0 \\ 0 & \frac{T_2}{v_2} \end{array} \right) 
\label{eq:pseudomass}
\\
{\cal M}_C &=& \left( \frac{m_{12}^2}{v_1 v_2} - \frac{\lambda_4 +
    \lambda_5}{2} \right) \left( \begin{array}{cc} v_2^2 & - v_1 v_2
    \\ -v_1 v_2 & v_1^2 \end{array} \right) +
\left( \begin{array}{cc} \frac{T_1}{v_1} & 0 \\ 0 &
    \frac{T_2}{v_2} \end{array} \right) \;,
\label{eq:chargedmass}
\eeq
where we introduced the abbreviation 
\beq
\lambda_{345} \equiv \lambda_3 + \lambda_4 + \lambda_5 \;.
\eeq
In Eqs.~(\ref{eq:scalarmass})-(\ref{eq:chargedmass}) we explicitly
kept the tadpole parameters $T_1$ and $T_2$, which vanish at tree
level, in order to ensure the correct treatment of the minimum
conditions beyond leading order (LO).
The diagonal mass matrices of the physical states can be obtained  
by performing the following orthogonal transformations
\beq
\left( \begin{array}{c} \rho_1 \\ \rho_2 \end{array} \right) &=&
R(\alpha) \left( \begin{array}{c} H \\ h \end{array} \right)  \; , \label{eq:diagHh} \\
\left( \begin{array}{c} \eta_1 \\ \eta_2 \end{array} \right) &=&
R(\beta) \left( \begin{array}{c} G^0 \\ A \end{array} \right)  \;
, \label{eq:diagGA} \\
\left( \begin{array}{c} \phi_1^\pm \\ \phi^\pm_2 \end{array} \right) &=&
R(\beta) \left( \begin{array}{c} G^\pm \\ H^\pm \end{array}
\right) \label{eq:diagGHpm}
\;, 
\eeq
which lead to the physical Higgs states, a neutral light CP-even, $h$, a neutral heavy
CP-even, $H$, a neutral CP-odd, $A$, and two charged Higgs bosons, 
$H^\pm$. The also obtained massless pseudo-Nambu-Goldstone bosons
$G^\pm$ and $G^0$ form the longitudinal components of the massive gauge
bosons, the charged $W^\pm$ and the $Z$ boson, respectively. In terms
of the mixing angles $\vartheta = \alpha$ and $\beta$, respectively,
the rotation matrices read
\beq
R(\vartheta) = \left( \begin{array}{cc} \cos \vartheta & - \sin
    \vartheta \\ \sin \vartheta & \cos \vartheta \end{array} \right) \;.
\eeq
The mixing angle $\beta$ can be expressed through the ratio of the two VEVs,
\beq
\tan \beta = \frac{v_2}{v_1} \;, \label{eq:tanbetadef}
\eeq
with the phenomenological constraint $v_1^2 + v_2^2 = v^2 \approx (246
\mbox{ GeV})^2$. The mixing angle $\alpha$ on the other hand can be
parametrized in terms of the entries $({\cal M}_S)_{ij}$ ($i,j=1,2$) of the
CP-even scalar mass matrix as
\beq
\tan 2\alpha = \frac{2 ({\cal M}_S)_{12}}{({\cal M}_S)_{11}-({\cal M}_S)_{22}} \;.
\eeq 
Introducing the abbreviation 
\beq
M^2 \equiv \frac{m_{12}^2}{s_\beta c_\beta}
\eeq
and the short-hand notation $s_x \equiv \sin x$ etc., we have
\cite{Kanemura:2004mg} 
\beq
\tan 2\alpha = \frac{s_{2\beta} (M^2- \lambda_{345} v^2)}{c_\beta^2
  (M^2-\lambda_1 v^2) -s_\beta^2 (M^2-\lambda_2 v^2)}
\;.  \label{eq:alphadef}
\eeq
After diagonalization the physical masses are given by
\beq
m_{h,H}^2 &=& \frac{1}{2} \left[ ({\cal M}_S)_{11} + ({\cal M}_S)_{22} \mp
\sqrt{\left(({\cal M}_S)_{11} - ({\cal M}_S)_{22}\right)^2 + 4 (({\cal
    M}_S)_{12})^2} \right]  \\
m_A^2 &=& M^2 -\lambda_5 v^2 \\
m_{H^\pm}^2 &=& M^2 - \frac{\lambda_4+\lambda_5}{2} v^2 \;.
\eeq
Note, in particular, that the masses of the heavier Higgs bosons,
$\phi_{\text{heavy}}=H, A$ and $H^\pm$, take
the form \cite{Kanemura:2004mg}
\beq
m_{\phi_{\text{heavy}}}^2 = c_{\phi_{\text{heavy}}}^2 M^2 +
f(\lambda_i) \, v^2 + {\cal O} (v^4/M^2) \;, \label{eq:massrels}
\eeq
where $f(\lambda_i)$ denotes a linear combination of $\lambda_1 -
\lambda_5$. The coefficient $c_{\phi_{\text{heavy}}}$ is given by
\beq
c_{\phi_{\text{heavy}}} = \left\{ \begin{array}{ll} 1 & \mbox{for } \;
    \phi_{\text{heavy}} = A, H^\pm \\
    s_{\beta-\alpha} & \mbox{for } \; \phi_{{\text{heavy}}} = H \end{array}
  \right. \;.
\eeq
There are two interesting limits that will play an important role in the relative
size of the NLO corrections. 
For $c_{\phi_{\text{heavy}}}^2 M^2 \gg f(\lambda_i) \, v^2$ we are in the {\it decoupling
limit}. {In the opposite case, if $M^2 \lsim f(\lambda_i) \, v^2$ and
the Higgs boson masses are large, we are in the {\it strong coupling
  regime}, as we then need the coupling strengths to
be significant. Both regimes will be investigated in detail in the
numerical analysis. \s 

For the parametrization of the Higgs potential $V$ there are 
various possibilities to choose the set of independent
parameters. Our guideline is given by the wish to relate the parameters to as many
physical quantities as possible. Thus we express the VEV $v$ in terms of
the physical gauge boson masses $M_W$ and $M_Z$ and the electric
charge $e$, and replace $m_{11}^2$ and $m_{22}^2$ by the tadpole
parameters $T_1$ and $T_2$. Later, we will also choose the renormalization
through Higgs decays. For this we need the fermion masses $m_f$. Our set of
independent parameters is then given by the Higgs boson masses,
the tadpole parameters, the two mixing angles, the soft breaking
parameter, the massive gauge boson masses, the electric
charge and the fermion masses: 
\beq
\mbox{\underline{Input parameters:}} \quad
m_h,\; m_H,\; m_A,\; m_{H^\pm},\; T_1,\; T_2,\; \alpha,\;
\tan\beta,\; m_{12}^2, \; M_W^2, \; M_Z^2,\; e , \; m_f \;.
\eeq

\section{Renormalization \label{sec:renormcond}}
\setcounter{equation}{0}
The one-loop computation of our sample Higgs-to-Higgs decay process 
\beq
H \to hh \;,
\eeq
encounters ultraviolet (UV) divergences. These are cancelled by the
renormalization of the parameters and wave functions involved in the
process. In particular, the process requires the renormalization of
the gauge sector and the Higgs sector of the 2HDM. In \cite{Krause:2016oke} we
proposed several renormalization schemes for the mixing angles
$\alpha$ and $\beta$, among these also the process-dependent
renormalization through the decays $H \to \tau \tau$ and $A \to \tau
\tau$. These processes additionally require the renormalization of the
fermion sector. Here, we first briefly repeat the basic features of our
chosen renormalization conditions that have been described in
\cite{Krause:2016oke}, with emphasis on the renormalization of the
mixing angles. For further details, we refer the reader to
\cite{Krause:2016oke}. We then present the renormalization
of the soft breaking parameter $m_{12}^2$, which is required in the
loop-corrected Higgs-to-Higgs decays. \s

For the renormalization, the bare parameters $p_0$ involved in the process
have to be replaced by the renormalized ones, $p$, and the corresponding
counterterms $\delta p$,
\beq
p_0 = p + \delta p \;.
\eeq
Additionally the fields $\Psi$ are renormalized by their field
renormalization constants $\delta Z_\Psi$ as
\beq
\Psi_0 = \sqrt{Z_\Psi} \Psi \;,
\eeq
where $\Psi$ generically stands for scalar, vector and fermion
fields. Note, that $Z_{\Psi}$ is a matrix in case of
  mixing fields. All Higgs bosons, gauge bosons and fermions are
renormalized 
on-shell (OS). The electric charge, which enters the weak gauge couplings,
is defined to be the full electron-positron photon coupling for
OS particles in the Thomson limit. Note, that we will use the
fine-structure constant at the $Z$ boson mass, $\alpha (M_Z^2)$, as
input in order to avoid large logarithms due to light fermions $f \ne
t$. The renormalization conditions for the tadpoles are chosen such
that the correct vacuum is reproduced at one-loop order which implies
\beq
\delta T_i = T_i \;, \qquad i=1,2 \;,
\eeq
where the $T_i$ stand for the contributions from the genuine Higgs
boson tadpole graphs in the gauge basis. 

\subsection{Renormalization of the Mixing Angles}
In  \cite{Krause:2016oke} we discussed in great detail the
renormalization of the mixing angles $\alpha$ and $\beta$. In
particular, schemes used in the literature before were shown to lead
to gauge-dependent decay amplitudes. This is based on the fact that
the standard treatment of the tadpoles, the {\it standard tadpole} scheme, leads to
gauge-dependent counterterms for the masses and mixing angles. In
particular, a gauge-independent decay amplitude can then only be obtained
through a physical, {\it e.g.}~a process-dependent, definition of the
angular counterterms. In the standard tadpole scheme the correct
vacuum at higher orders is given by the VEV\footnote{In the 2HDM we
  have two VEVs, which are related, however, due to the
  requirement of ensuring unitarity of the scattering
  amplitudes.}  
that is derived from the
gauge-dependent loop-corrected Higgs potential, and is therefore also
gauge dependent. Consequently, all bare quantities and counterterms
given in terms of the VEV become gauge dependent as well. In the {\it
alternative tadpole} scheme \cite{Fleischer:1980ub}, the bare quantities are not
gauge dependent, as they are expressed in terms of the tree-level
VEV, which is gauge independent. The correct minimum at higher
orders is reproduced by shifting the VEV. The shift affects the
counterterms but not the bare quantities. With the exception of the wave
function renormalization constants, the counterterms are gauge
independent in the alternative tadpole scheme. In practice the change
from the standard to 
the alternative tadpole scheme, also referred to as {\it standard} and
{\it tadpole} scheme, respectively, requires the following modifications:
\begin{itemize}
\item {\it Self-energies:} The self-energies in the wave function
  renormalization constants and counterterms have to be changed to
  contain additional tadpole contributions: $\Sigma(p^2) \to
  \Sigma^{\text{tad}} (p^2)$.
\item {\it Tadpole counterterms:} In turn, the tadpole counterterms do
  not appear any more in the scalar sector: $\delta T_{\phi_i \phi_j} \to
  0$.
\item {\it Vertex corrections:} In the virtual corrections additional
  tadpole contributions have be taken into account if the
  extension of the corresponding coupling by an external CP-even Higgs
  boson $h,H$, which carries the tadpole, exists. 
\end{itemize}
For all details, we refer the reader to Appendix A of
\cite{Krause:2016oke}. \s

In \cite{Krause:2016oke} the {\it tadpole-pinched} scheme was
introduced as a manifestly
gauge-independent renormalization scheme for the angular counterterms. 
It relies on the use of the alternative tadpole scheme together with the
modified Higgs self-energies defined by means of the
pinch technique 
\cite{Cornwall:1989gv,Papavassiliou:1989zd,Degrassi:1992ue,Papavassiliou:1994pr,Watson:1994tn,Binosi:2002ft,Binosi:2009qm}.\footnote{For
a discussion of the pinch technique, see \cite{Papavassiliou:1995fq,
Papavassiliou:1995gs, Papavassiliou:1996zn, Pilaftsis:1996fh,
Papavassiliou:1997fn, Papavassiliou:1997pb} and also
\cite{Binosi:2004qe,Binosi:2009qm} for a comparison with the
background field method
\cite{Abbott:1980hw,Abbott:1981ke,KlubergStern:1974xv,KlubergStern:1975hc,Boulware:1980av,Hart:1984jy,Denner:1994xt,Denner:1994nn}.}
The 
angular counterterms are obtained in terms of the 
pinched self-energies $\overline{\Sigma} (p^2)$, where $p^2$ denotes the
four-momenta squared at which they are evaluated. Note that they have
to be evaluated in the tadpole scheme and can be related to the
tadpole self-energies in the Feynman gauge through
\beq
\overline{\Sigma} (p^2) = \left.\Sigma^{\text{tad}}
  (p^2)\right|_{\xi=1} + \Sigma^{\text{add}} (p^2) \;.
\eeq
Here $\xi$ represents the gauge fixing parameters $\xi_Z,$ $\xi_W$ and
$\xi_\gamma$ of the $R_\xi$ gauge. For the renormalization of the
mixing angle $\beta$ the pseudoscalar or the charged sector can
be used, leading to different counterterm definitions. We will use two
different definitions, specified below. We will furthermore apply two
different tadpole-pinched schemes which differ by their choice of the
renormalization scale: \s 

\noindent 
\underline{\it On-shell tadpole-pinched scheme:}
The renormalization scale is chosen to be the on-shell scale in the
appearing self-energies. Applying
  \cite{Espinosa:2002cd}, the angular counterterms are given by
\beq
\delta \alpha &=& \frac{\mbox{Re}\left(\left[\Sigma_{Hh}^{\text{tad}} (m_H^2) +
  \Sigma^{\text{tad}}_{Hh} (m_h^2) \right]_{\xi=1} +
\Sigma^{\text{add}}_{Hh} (m_H^2) + \Sigma^{\text{add}}_{Hh} (m_h^2)
\right)}{2 (m_H^2 - m_h^2)}  
\label{eq:posalpha}
\\
\delta \beta^{(1)} &=& - \frac{\mbox{Re}\left(\left[\Sigma_{G^\pm
        H^\pm}^{\text{tad}} (0) + 
  \Sigma^{\text{tad}}_{G^\pm H^\pm} (m_{H^\pm}^2) \right]_{\xi=1} +
  \Sigma^{\text{add}}_{G^\pm H^\pm} (0)
  + \Sigma^{\text{add}}_{G^\pm H^\pm} (m_{H^\pm}^2)\right)}{2 m_{H^\pm}^2} 
\label{eq:posbeta1} \\
\delta \beta^{(2)} &=& - \frac{\mbox{Re}\left(\left[\Sigma_{G^0 A}^{\text{tad}} (0) +
  \Sigma^{\text{tad}}_{G^0 A} (m_A^2) \right]_{\xi=1} +
  \Sigma^{\text{add}}_{G^0 A} (0)
  + \Sigma^{\text{add}}_{G^0 A} (m_A^2)\right)}{2 m_A^2} 
\;. \label{eq:posbeta2}
\eeq
The additional contributions read (see also \cite{Espinosa:2002cd} for
the CP-even case in the MSSM),
\beq
\Sigma^{\text{add}}_{Hh} (p^2) &=& \frac{g^2 s_{\beta -\alpha}
  c_{\beta-\alpha}}{32 \pi^2 c_W^2} \left( p^2 - \frac{m_H^2 +
    m_h^2}{2} \right) \Big\{ B_0 (p^2; m_Z^2, m_A^2) -B_0 (p^2; m_Z^2,
m_Z^2) \nonumber
\\
&& + 2 c_W^2 \left[ B_0 (p^2; m_W^2, m_{H^\pm}^2) -B_0 (p^2; m_W^2,
  m_W^2) \right] \Big\}
\label{eq:sigaddhh} \\
\Sigma^{\text{add}}_{G^0 A} (p^2) &=& \frac{g^2 s_{\beta -\alpha}
  c_{\beta-\alpha}}{32 \pi^2 c_W^2} \left( p^2 - \frac{m_A^2}{2}
\right) \left[ B_0 (p^2; m_Z^2,m_H^2) - B_0 (p^2; m_Z^2, m_h^2)
\right] \label{eq:sigaddga} \\
\Sigma^{\text{add}}_{G^\pm H^\pm} (p^2) &=& \frac{g^2 s_{\beta -\alpha}
  c_{\beta-\alpha}}{16 \pi^2} \left( p^2 - \frac{m_{H^\pm}^2}{2}
\right) \left[ B_0 (p^2; m_W^2,m_H^2) - B_0 (p^2; m_W^2, m_h^2)
\right] \;, \label{eq:sigaddghpm} 
\eeq 
where $B_0$ is the scalar two-point function
\cite{'tHooft:1978xw,Passarino:1978jh} and $c_W$ refers to the cosine
of the Weinberg angle $\theta_W$. \s

\noindent
\underline{\it $p_\star$ tadpole-pinched scheme:}
In this scheme the self-energies are evaluated at the average of the
particle momenta squared \cite{Espinosa:2002cd},
\beq
p_\star^2 = \frac{m_{\phi_1}^2 + m_{\phi_2}^2}{2} \;,
\eeq
with $(\phi_1,\phi_2)=(H,h)$, $(G^\pm,H^\pm)$ and $(G^0,A)$,
respectively. The additional contributions then obviously vanish and
the angular counterterms simplify to
\beq
\delta \alpha &=& \frac{\mbox{Re} \left[\overline{\Sigma}_{Hh} \left(
      \frac{m_h^2 + 
      m_H^2}{2} \right) \right]}{m_H^2-m_h^2} \label{eq:delalphstar} \\
\delta \beta^{(1)} &=& - \frac{\mbox{Re}
  \left[\overline{\Sigma}_{G^\pm H^\pm} \left( 
    \frac{m_{H^\pm}^2}{2} \right)\right]}{m_{H^\pm}^2} \label{eq:delbet1star}
\\
\delta \beta^{(2)} &=& - \frac{\mbox{Re} \left[\overline{\Sigma}_{G^0 A}
  \left( \frac{m_A^2}{2} \right) \right]}{m_A^2} \label{eq:delbet2star} \;.
\eeq

\noindent
\underline{\it Process-dependent renormalization:}
We also apply a process-dependent renormalization of the mixing
angles. The angular counterterm $\delta\beta$ is obtained from the
requirement that the loop-corrected Higgs decay $A\to \tau\tau$
including only the weak corrections is equal to the LO
width\footnote{See \cite{Freitas:2002um}, for a discussion on the renormalization
  of $\tan\beta$ within the MSSM and the application of the
  process-dependent scheme.},
\beq
\Gamma^{\text{LO}} (A\to \tau\tau) \stackrel{!}{=}
\Gamma^{\text{NLO}}_{\text{weak}} (A\to \tau\tau) \;.
\eeq
The counterterm $\delta\alpha$ is
obtained by applying the same condition, but on the $H \to \tau\tau$
decay,
\beq
\Gamma^{\text{LO}} (H\to \tau\tau) \stackrel{!}{=}
\Gamma^{\text{NLO}}_{\text{weak}} (H\to \tau\tau) \;.
\eeq
The process-dependent renormalization leads to gauge-dependent
angular counterterms if the standard tadpole scheme is applied. The
angular counterterms are manifestly gauge independent, on the other
hand, in case the alternative tadpole scheme is used. 

\subsection{Renormalization of $m_{12}^2$}
For the renormalization of the soft $\mathbb{Z}_2$ breaking parameter
$m_{12}^2$ the bare parameter is replaced by the renormalized one and
its counterterm,
\beq
(m_{12}^2)_0 = m_{12}^2 + \delta m_{12}^2 \;.
\eeq
We will apply two different renormalization schemes. \s

\noindent
\underline{\it Modified Minimal Subtraction Scheme:} In the modified minimal
subtraction ($\overline{\mbox{MS}}$) scheme\footnote{We did not apply the
  $\overline{\mbox{MS}}$ scheme to the 
  renormalization of the mixing angles, as it  leads to one-loop
  corrections of the decay widths that are orders of magnitude larger
  than in the other schemes. This was checked in \cite{masterlorenz}
  for a large set of allowed 2HDM scenarios. The reason is that in general
  the wave function renormalization constants introduce large finite
  contributions to the one-loop amplitudes, which need to be 
cancelled by the finite parts of the angular counterterms, a
cancellation that does not take place any more in the $\overline{\mbox{MS}}$
scheme.} the counterterm $\delta m_{12}^2$ is chosen such that it
cancels all residual terms of the amplitude, which are proportional to  
\beq
\Delta = \frac{1}{\epsilon} - \gamma_E + \ln (4\pi) \;,
\eeq
where $\gamma_E$ denotes the Euler-Mascheroni constant. 
These terms obviously contain the remaining UV divergences given as
poles in $\epsilon$ plus additional finite constants that appear
universally in all loop integrals \cite{Olness:2008ty}. The
renormalization of $\delta m_{12}^2$ in this scheme is hence given by
\beq
\delta m_{12}^2 = \delta m_{12}^2 (\Delta)|_{\overline{\text{MS}}} \;,
\eeq
where the right-hand side of the equation symbolically denotes all
terms proportional to $\Delta$ that are necessary to cancel the
$\Delta$ dependence of the remainder of the amplitude. 
\s

\noindent
\underline{\it Process-dependent renormalization:} A more physical
definition of the counterterm is provided by the renormalization
through a physical process. As $m_{12}^2$ only appears in the
couplings between Higgs bosons, the simplest processes that can be
chosen to fix the counterterm are given by the on-shell decays 
\beq
H &\to& hh \\
H &\to& H^+ H^- \\
h &\to& AA \\
H &\to& AA \;.
\eeq
As the scalar $h$ is identified with the 125 GeV Higgs boson the decay
$h\to H^+ H^-$ is kinematically not possible, since we restrict the
charged Higgs mass to $m_{H^\pm} > m_h$, see {\it
  e.g.}~\cite{Misiak:2015xwa} for a type 
II 2HDM.\footnote{The 2HDM also allows for scenarios with the second
  lightest Higgs boson $H$ being the SM-like resonance. This kinematic
  set-up would worsen the situation here, however.} We will compute the loop
corrections to the decay $H \to hh$ in order to study the impact of
the various renormalization schemes, so that this process cannot be
used for the determination of $\delta m_{12}^2$. With $H^\pm$ masses
above 480~GeV in the type II 2HDM \cite{Misiak:2015xwa}, which we will
choose for the numerical analysis, the decay $H\to H^+ H^-$ would
require very heavy $H$ bosons, so that we do not consider this process either.
The OS process $h\to AA$ is kinematically very restricted as it
requires pseudoscalars $A$ 
with masses below 125 GeV/2 that additionally have escaped detection at
collider experiments so far. Although such scenarios are possible in
principle, they are very rare, and the measurement of the decay is
challenging. This leaves us with the process $H\to AA$ as the least
restrictive one to fix the counterterm of $\delta m_{12}^2$. \s

Note, that $\delta m_{12}^2$ in both schemes is gauge independent
irrespective of the chosen tadpole scheme. Being a parameter of
the original 2HDM Higgs potential before EWSB, it is not related to the VEV and hence
cannot encounter any gauge dependences arising from the treatment of the
VEV at higher orders. 

\section{Decay Widths at Electroweak One-Loop Order \label{sec:nlocalc}}
We will present here the details for the computation of the
electroweak one-loop corrections to the Higgs-to-Higgs decay widths 
\beq
H &\to& hh  \qquad \mbox{ and}\\
H &\to& AA \;.
\eeq
The first process will be used to study numerically the impact of the
various renormalization schemes that we propose on the NLO
corrections. The second process serves for a process-dependent
definition of the counterterm $\delta m_{12}^2$.  

\subsection{Electroweak One-Loop Corrections to $H\to hh$ \label{subsec:Htohh}}
The heavy Higgs decay into a pair of SM-like Higgs bosons,
\beq
H \to hh \;,
\eeq
depends through the trilinear Higgs self-coupling
\beq
\lambda_{Hhh} \equiv g \cdot g_{Hhh} = g \frac{-c_{\beta-\alpha}}{2 M_W
  s_{2\beta}} \left( s_{2\alpha} (2m_h^2 + m_H^2) -
  \frac{m_{12}^2}{s_\beta c_\beta} (3 s_{2\alpha} -s_{2\beta} )
\right)
\label{eq:gheavyhh}
\eeq
not only on the mixing angles $\alpha$ and $\beta$ but also on
$m_{12}^2$. The LO decay width is given by
\beq 
\Gamma^{\text{LO}} (H \to hh) = \frac{G_F M_W^2 m_H g_{Hhh}^2}{4
  \sqrt{2} \pi} \sqrt{1-\frac{4m_h^2}{m_H^2}} \;,
\eeq
where $G_F$ denotes the Fermi constant.
The NLO decay width can be written as the sum of the LO width and the
one-loop corrected decay width $\Gamma^{(1)}$, 
\beq
\Gamma^{\text{NLO}} = \Gamma^{\text{LO}} + \Gamma^{(1)} \;.
\eeq
The one-loop correction $\Gamma^{(1)}$ is obtained from the
interference of the LO decay amplitude with the one at NLO. The
contributions to the NLO decay amplitude are given by the virtual
corrections and the counterterm diagrams. The virtual corrections
consist of the pure vertex corrections, shown in
Fig.~\ref{fig:vertexcorrs}, and the corrections to the external
legs. The vertex corrections comprise the one-particle irreducible
(1PI) diagrams given by the triangle diagrams with fermions, scalars, 
ghosts and gauge bosons in the loops and the diagrams involving
four-particle vertices. 
\begin{figure}[t!]
  	\centering
  	\includegraphics[width=400pt, trim = 0mm 5mm 0mm 4mm, clip]{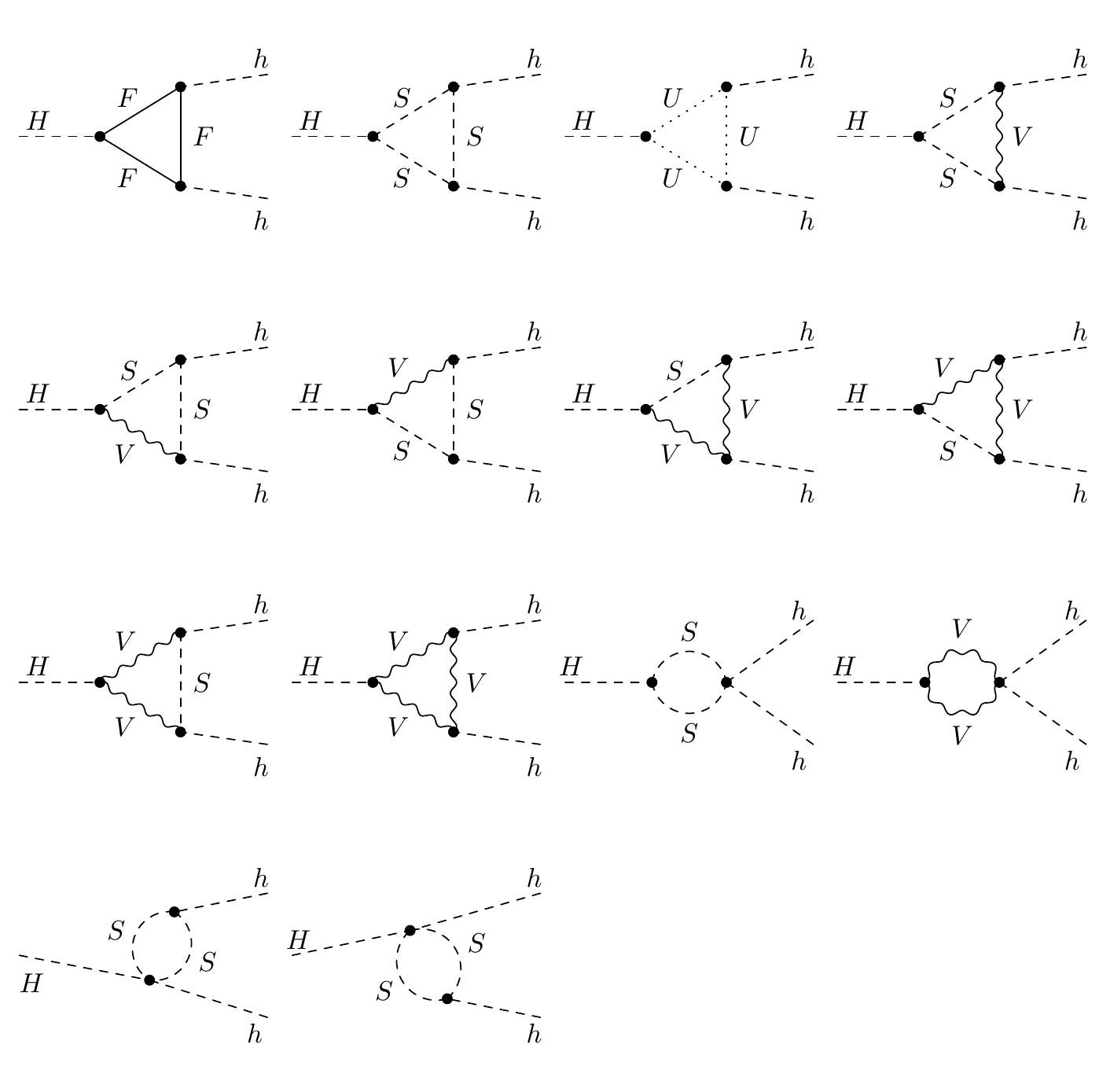}
    \caption{Generic diagrams contributing to the vertex corrections
      in $H \to hh$ with fermions $F$, scalar bosons $S$, 
      ghosts $U$ and gauge bosons $V$ in the loops.}
\label{fig:vertexcorrs}
\end{figure} 
The external leg corrections consist of off-diagonal and diagonal
field mixing contributions $hH, Hh, HH$ and $hh$, which all vanish due
to the OS renormalization conditions of the external fields. The
counterterm diagrams are shown in Fig.~\ref{fig:ct}. They are given by
all possible counterterm insertions on the external legs and the
genuine vertex counterterm. 
\begin{figure}[t!]
  	\centering
\includegraphics[width=400pt, trim = 0mm 5mm 0mm 4mm, clip]{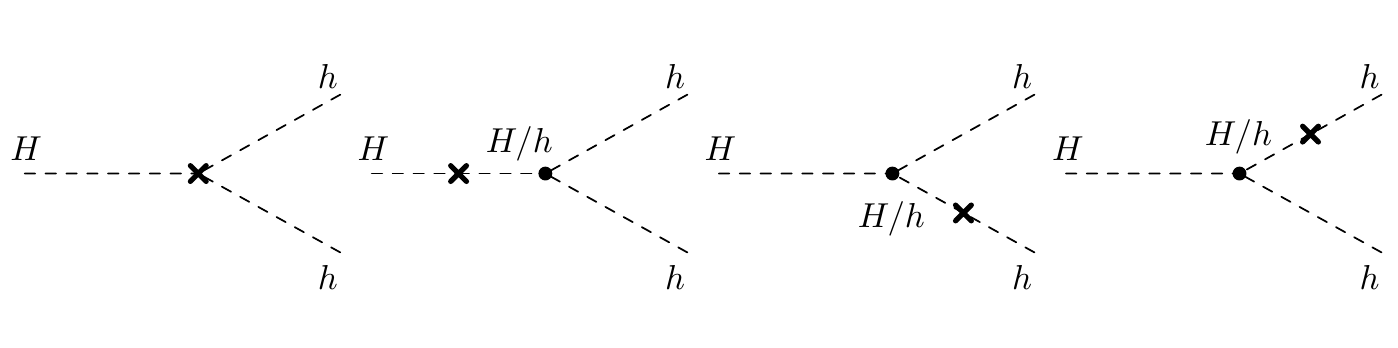}
    \caption{Counterterm diagrams contributing to the NLO decay $H\to
      hh$.}
\label{fig:ct}
\end{figure} 
For the correct derivation of the symmetry factors associated with the
various counterterm contributions we start from the bare Lagrangian
describing the involved trilinear Higgs self-interactions. In terms of
the coupling factors
\beq
g_{hhh} &=& \frac{3}{2 M_W s_{2\beta}} \left( \frac{2 m_{12}^2}{s_\beta
    c_\beta} c_{\alpha+\beta} c_{\beta-\alpha}^2 - m_h^2
  (2c_{\alpha+\beta} + s_{2\alpha} s_{\beta-\alpha} ) \right) \\
g_{HHh} &=& \frac{s_{\beta-\alpha}}{2M_W s_{2\beta}} \left( -
  \frac{m_{12}^2}{s_\beta c_\beta} (3 s_{2\alpha} + s_{2\beta} )
+  s_{2\alpha} (m_h^2 + 2m_H^2) \right)
\eeq
and $g_{Hhh}$ defined in Eq.~(\ref{eq:gheavyhh}) it reads
\beq
{\cal L}_{\text{int}}^{Hhh} &=& g \left[ \frac{- g_{hhh}}{3!} h_0 h_0 h_0 
- \frac{g_{Hhh}}{2!} H_0 h_0 h_0 - \frac{g_{HHh}}{2!} H_0 H_0 h_0
\right]
\;,
\eeq
where $h_0$ and $H_0$ denote the bare fields. At NLO we obtain in
terms of the renormalized fields $h$ and $H$,
\beq 
{\cal L}_{\text{int}}^{Hhh} &\stackrel{\text{NLO}}{\approx}& g \left[
  \frac{- g_{hhh}}{3!} 
  \frac{3\delta Z_{hH}}{2} - \frac{g_{Hhh}}{2!} \left( \delta Z_{hh} +
    \frac{\delta Z_{HH}}{2} \right) - \frac{g_{HHh}}{2!} \delta Z_{Hh}
\right] Hhh \;,
\eeq
where the $\delta Z$'s denote the wave function renormalization
constants. 
The Feynman rule $\lambda_{\text{CT,WR}}^{Hhh}$ for this counterterm
contribution from the wave function renormalization is derived by
applying the functional derivatives with respect to the external
renormalized fields,
\beq
\lambda_{\text{CT,WR}}^{Hhh} &=& i \frac{\delta}{i \delta H}
\frac{\delta}{i \delta h} \frac{\delta}{i \delta h} {\cal
  L}_{\text{int}}^{Hhh} \;.
\eeq
Adding the genuine vertex counterterm $\delta (g \cdot g_{Hhh})$
we have for the counterterm amplitude
\beq
{\cal M}_{Hhh}^{\text{CT}} = g\left[ g_{hhh} \frac{\delta Z_{hH}}{2} +
g_{Hhh} \left( \delta Z_{hh} + \frac{\delta Z_{HH}}{2} \right) +
g_{HHh} \delta Z_{Hh} + \frac{1}{g} \delta (g \cdot g_{Hhh}) \right] \;.
\eeq
The genuine vertex counterterm at NLO is given by
\beq
\delta ( g \cdot g_{Hhh}) &=& g \Big\{ g_{Hhh} \left( \frac{\delta
      g}{g} - \frac{\delta M_W}{M_W} \right)  \nonumber \\
&& \; + \left(
    \frac{-c_{\beta-\alpha}}{M_W s_{2\beta}} \right)
    \left[\frac{s_{2\alpha}}{2} (2 \delta m_h^2 + \delta m_H^2) -
      \left( \frac{3s_{2\alpha}-s_{2\beta}}{s_{2\beta}} \right)\delta
        m_{12}^2 \right] \nonumber \\
&& \; + \left[ g_{Hhh} \left(-t_{\beta-\alpha} - \frac{2}{t_{2\beta}}
  \right) - \frac{m_{12}^2}{M_W} \left(
    \frac{c_{\beta-\alpha}}{s_{2\beta}^2} \right) \frac{6
    s_{2\alpha}}{t_{2\beta}} \right] \delta\beta \nonumber \\
&& \; + \left[ g_{Hhh} t_{\beta-\alpha} -
  \frac{2m_h^2 + m_H^2 - 3 m_{12}^2/(s_\beta c_\beta)}{M_W} 
  \frac{c_{\beta-\alpha} c_{2\alpha}}{s_{2\beta}} \right] \delta \alpha
\Big\} \;.
\eeq
The NLO corrections factorize from the LO amplitude so that the
one-loop corrected decay width can be cast into the form
\beq
\Gamma^{\text{NLO}} (H\to hh) = \Gamma^{\text{LO}} \left[ 1+
  \Delta^{\text{virt}}_{Hhh} + \Delta^{\text{ct}}_{Hhh} \right] \;, \label{eq:nlodecay}
\eeq
with $\Delta^{\text{ct}}_{Hhh}$ given by
\beq
\Delta^{\text{ct}} = \frac{2 {\cal M}_{Hhh}^{\text{CT}}}{g \cdot g_{Hhh}} \;.
\eeq
The expression $\Delta^{\text{virt}}_{Hhh}$ is quite lengthy so that we do
not display it explicitly here. \s

\begin{figure}[tb]
  	\centering
  	\includegraphics[width=\linewidth, trim = 0mm 5mm 0mm 1mm, clip]{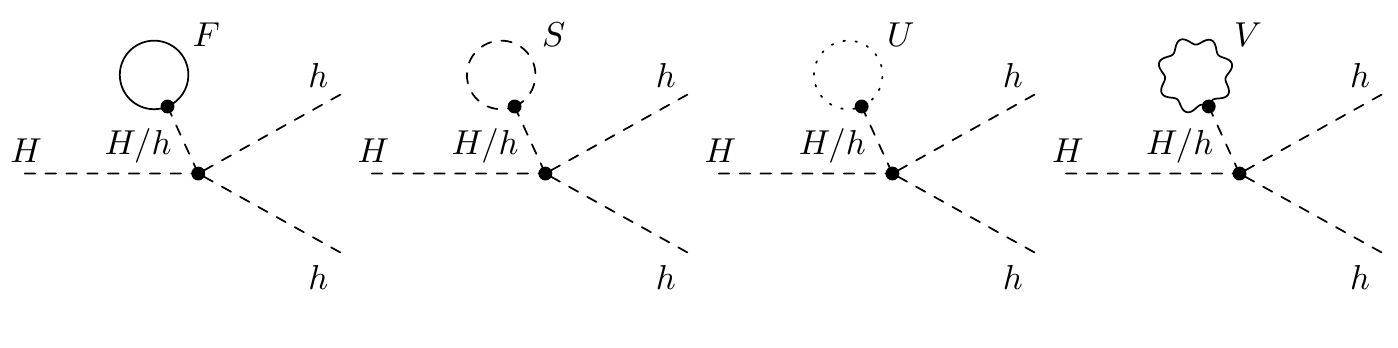}
    \caption{Additional vertex diagrams in the alternative
        tadpole scheme contributing to the decay $H\to
        hh$.}
    \label{fig:tadpcontrs}
\end{figure}
In case the alternative tadpole scheme is applied, additional
diagrams have to be included in the virtual corrections. They are
depicted in Fig.~\ref{fig:tadpcontrs} and involve quartic Higgs
self-couplings where the additionally attached Higgs to the original
trilinear vertex is connected to a tadpole diagram. 
The inclusion of these additional diagrams in
  combination with the change of the mass, angular and wave function
  counterterms in the alternative tadpole scheme leaves the overall
  NLO decay width invariant, provided the angular counterterms are
  defined in a process dependent scheme.

\subsection{Electroweak One-Loop Corrections to $H\to AA$}
We use the decay of the heavy scalar $H$ into a pair of pseudoscalars
$A$, 
\beq
H \to AA \;,
\eeq
for a process-dependent renormalization of $m_{12}^2$. The leading
order decay width depends through the trilinear coupling
\beq
\lambda_{HAA} \equiv g \cdot g_{HAA} = - \frac{g}{2M_W} \left[ 
c_{\beta-\alpha} (2m_A^2 - m_H^2) +
\frac{s_{\alpha+\beta}}{s_{2\beta}} \left( 2 m_H^2 - \frac{2
    m_{12}^2}{s_\beta c_\beta} \right) \right] \label{eq:HAAcoup}
\eeq
besides on the mixing angles $\alpha$ and $\beta$ in particular on
$m_{12}^2$. The LO decay width is given by
\beq
\Gamma^{\text{LO}} (H \to AA) = \frac{G_F M_W^2 m_H g_{HAA}^2}{4
  \sqrt{2} \pi} \sqrt{1-\frac{4m_A^2}{m_H^2}} \;.
\eeq
The electroweak (EW) one-loop corrections consist of the virtual corrections and the
counterterm contributions which guarantee the UV-finiteness of the
decay amplitude. The virtual corrections, which comprise the corrections to
the external legs and the pure vertex corrections, are depicted in
Fig.~\ref{fig:virtualHAA}. The corrections to the external legs in
Fig.~\ref{fig:virtualHAA} (b), (c) and (d) vanish because of the OS
renormalization of the external fields. Diagrams (e) and (f) are zero
due to a Slavnov-Taylor identity \cite{Williams:2011bu}. 
\begin{figure}[tb]
  	\centering
  	\includegraphics[width=\linewidth, trim = 0mm 0mm 0mm 1mm, clip]{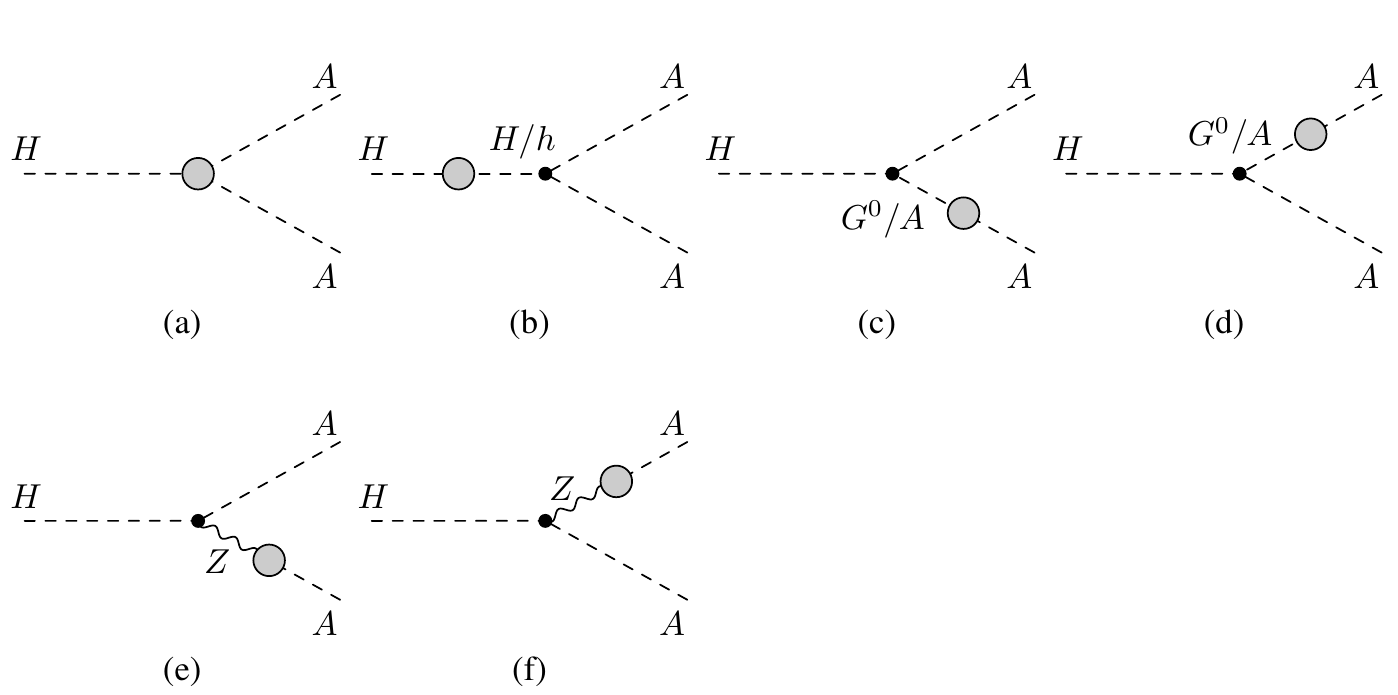}
    \caption{Generic diagrams contributing to the virtual corrections
      of $H\to AA$: vertex corrections (a) and corrections to the
      external legs (b)-(f).}
    \label{fig:virtualHAA}
\end{figure}
The 1PI diagrams of the vertex corrections are displayed in
Fig.~\ref{fig:vertexcorrsHAA}. They consist of the 1PI diagrams given by the 
triangle diagrams with fermions, scalars and gauge bosons in the loops
and by the diagrams containing four-particle vertices. 
\begin{figure}[tb]
  \centering
  \includegraphics[width=0.9\linewidth, trim = 0mm 5mm 0mm 5.9mm, clip]{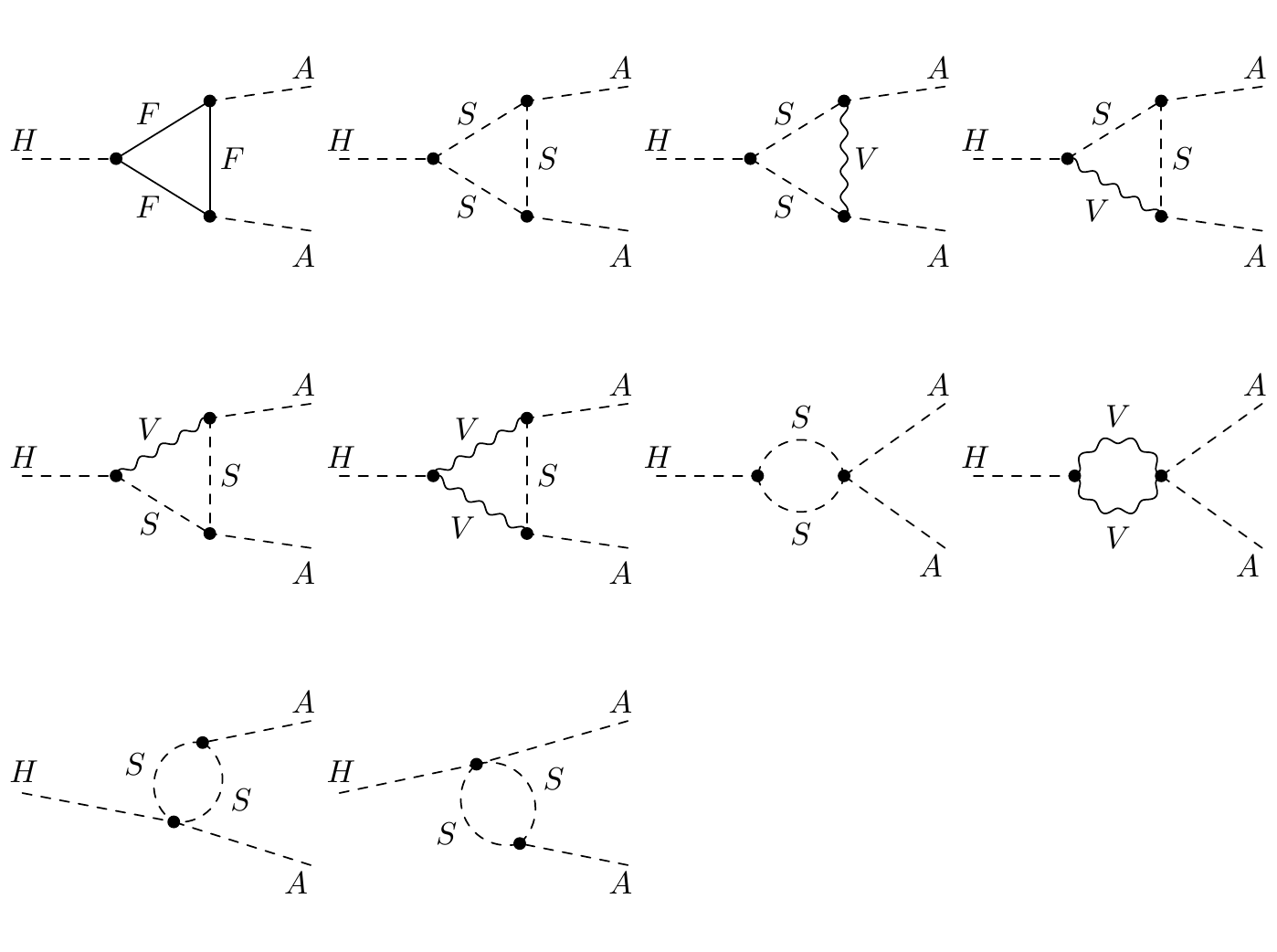}
\caption{Generic diagrams contributing to the vertex corrections
      in $H \to AA$ with fermions $F$, scalar bosons $S$
      and gauge bosons $V$ in the loops.}
\label{fig:vertexcorrsHAA}
\end{figure} 
The counterterm contributions are given by the genuine vertex
counterterm  and by the counterterm insertions on
the external legs, {\it cf.}~Fig.~\ref{fig:ctHAA}. For the derivation of
the latter we start from the bare Lagrangian involving the relevant
trilinear Higgs self-couplings. With the coupling factors
\beq
g_{hAA} &=& \frac{1}{2M_W} \left[ s_{\beta-\alpha} (2m_A^2 - m_h^2) +
  \frac{c_{\alpha+\beta}}{s_{2\beta}} \left( 2m_h^2 -
    \frac{2m_{12}^2}{s_\beta c_\beta} \right) \right] \\
g_{HG^0 A} &=& - \frac{s_{\beta-\alpha}}{2M_W} (m_A^2 - m_H^2) 
\eeq
and $g_{HAA}$ defined in Eq.~(\ref{eq:HAAcoup}) it reads in terms of
the bare fields denoted by the subscript $0$,
\beq
{\cal L}_{\text{int}}^{HAA} = g \left[ - \frac{g_{hAA}}{2!} h_0 A_0 A_0 -
  \frac{g_{HAA}}{2!} H_0 A_0 A_0 - g_{HG^0 A} H_0 G_0^0 A_0\right] \,.
\eeq
Replacing the bare fields by their renormalized ones and the
corresponding wave function renormalization constants, the NLO
expansion of the Lagrangian reads
\beq
{\cal L}_{\text{int}}^{HAA} \stackrel{\text{NLO}}{\approx} g \left[
- \frac{g_{hAA}}{2!} \frac{\delta Z_{hH}}{2} - \frac{g_{HAA}}{2!}
\left( \delta Z_{AA} +  \frac{\delta Z_{HH}}{2} \right) - g_{HG^0 A}
\frac{\delta Z_{G^0 A}}{2} \right] HAA \;.
\eeq
\begin{figure}[b!]
  	\centering
  	\includegraphics[width=400pt, trim = 0mm 5mm 0mm 4mm, clip]{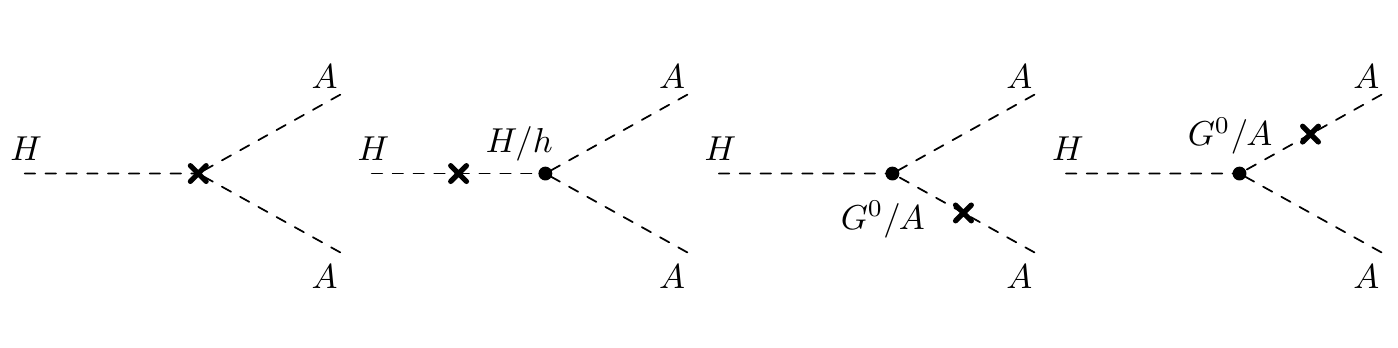}
    \caption{Counterterm diagrams contributing to the NLO decay $H\to
      AA$.}
\label{fig:ctHAA}
\end{figure} 
The Feynman rule $\lambda_{\text{CT,WR}}^{HAA}$ for the counterterm
contribution from the wave function renormalization is obtained by
performing the functional derivatives with respect to the external
renormalized fields,
\beq
\lambda_{\text{CT,WR}}^{HAA} &=& i \frac{\delta}{i \delta H}
\frac{\delta}{i \delta A} \frac{\delta}{i \delta A} {\cal
  L}_{\text{int}}^{HAA} \;.
\eeq 
Together with the genuine vertex counterterm $\delta (g \cdot
g_{HAA})$
\beq
\delta ( g \cdot g_{HAA}) &=& g \Big\{ g_{HAA} \left( \frac{\delta
      g}{g} - \frac{\delta M_W}{M_W} \right)  \nonumber \\
&& \hspace*{-1.5cm} - \frac{1}{2 M_W} 
    \left[ c_{\beta-\alpha} (2 \delta m_A^2-\delta m_H^2) +
      \frac{s_{\alpha+\beta}}{s_{2\beta}} \left( 2 \delta m_H^2 
 -\frac{4}{s_{2\beta}} \delta m_{12}^2 \right) \right]
\nonumber \\
&& \hspace*{-1.5cm} - \frac{1}{2 M_W} \left[ s_{\alpha-\beta} (2 m_A^2- m_H^2) +
  \frac{2(c_\alpha s_\beta^3 - s_\alpha c_\beta^3)}{s_{2\beta}^2}
  \left( 2 m_H^2 - \frac{4m_{12}^2}{s_{2\beta}} \right) + \frac{8
    c_{2\beta} s_{\alpha+\beta} m_{12}^2}{s_{2\beta}^3} \right] \delta
\beta \nonumber \\
&& \hspace*{-1.5cm} 
- \frac{1}{2 M_W} \left[ - s_{\alpha-\beta} (2m_A^2 -m_H^2) +
  \frac{c_{\alpha+\beta}}{s_{2\beta}} \left( 2m_H^2 - \frac{4
      m_{12}^2}{s_{2\beta}} \right) \right] \delta \alpha \Big\} \;. 
\eeq
we obtain for the counterterm amplitude
\beq
{\cal M}^{\text{CT}}_{HAA} = g \left[ g_{hAA} \frac{\delta Z_{hH}}{2}
  + g_{HAA} \left( \delta Z_{AA} +  \frac{\delta Z_{HH}}{2} \right) + g_{HG^0 A}
\delta Z_{G^0 A} + \frac{1}{g} \delta (g \cdot g_{HAA} )
\right] \;.
\eeq
The one-loop amplitude ${\cal M}_{HAA}^{\text{1loop}}$ of the decay
$H\to AA$ consists of the amplitude built from the vertex corrections ${\cal
  M}_{HAA}^{\text{VC}}$ and of the counterterm amplitude, 
\beq
{\cal M}_{HAA}^{\text{1loop}} = {\cal M}_{HAA}^{\text{VC}} + {\cal
  M}^{\text{CT}}_{HAA} \;.
\eeq
With the LO amplitude ${\cal M}_{HAA}^{\text{LO}}$ we then obtain the
NLO partial decay width as 
\beq
\Gamma^{\text{NLO}} = \Gamma^{\text{LO}} + \frac{m_H }{32 \pi} \sqrt{1-\frac{4
      m_A^2}{m_H^2}} \, 2 \, \mbox{Re} \left[ ({\cal 
    M}_{HAA}^{\text{LO}})^* {\cal M}_{HAA}^{\text{1loop}} \right] \;.
\eeq
The counterterm $\delta m_{12}^2$ is fixed by the process-dependent
renormalization condition
\beq
\Gamma^{\text{LO}} (H\to AA) \stackrel{!}{=} \Gamma^{\text{NLO}} (H\to
AA) \;.
\eeq
This leads to the counterterm definition 
\beq
\delta m_{12}^2 = -\frac{v s_{2\beta}^2}{4 s_{\alpha+\beta}} \mbox{Re} \left[  {\cal
    M}_{HAA}^{\text{VC}} + ({\cal M}^{\text{CT}}_{HAA})_{\delta
    m_{12}^2=0} \right] \;.
\eeq

The additional diagrams that must be taken into account when the
alternative tadpole scheme is a applied are displayed in
Fig.~\ref{fig:tadphtoaa}. Note that the overall NLO amplitude is 
invariant under a change of the tadpole schemes, provided the angular
counterterms are determined in a process-dependent way.
\begin{figure}[tb]
  	\centering
  	\includegraphics[width=\linewidth, trim = 0mm 5mm 0mm 1mm, clip]{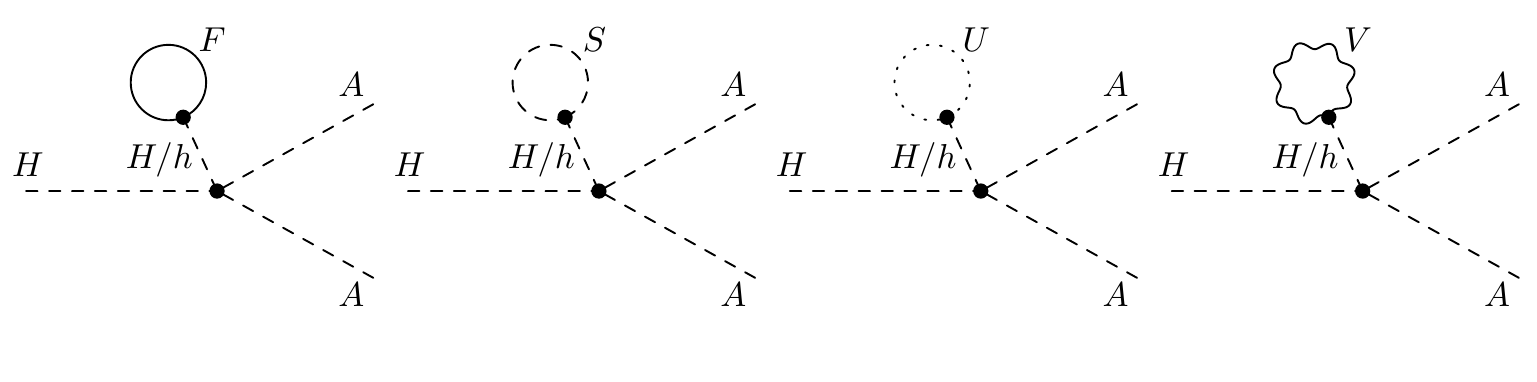}
    \caption{Additional vertex diagrams in the alternative
        tadpole scheme contributing to the decay $H\to
        AA$.}
    \label{fig:tadphtoaa}
\end{figure}

\subsection{Gauge (In)dependence of the NLO amplitude}
As the expressions for the vertex corrections and counterterms are
quite involved we limit our discussion here to a qualitative
level. The quantitative corroboration of our statements will be
presented in the numerical analysis. \s

In case the standard tadpole scheme is applied the computation of  
the NLO decay amplitude in the general $R_\xi$ gauge reveals that the
residual amplitude ${\cal M}$ with the counterterms 
\beq
\delta p \equiv \delta \alpha \;, \; \delta \beta
\mbox{ and } \delta m_{12}^2
\eeq 
set to zero exhibits a  UV-divergent gauge dependence,
\beq
\underline{\mbox {\it Standard tadpole scheme:}} \qquad 
{\cal M}_{H \to hh}|_{\text{NLO}, \xi, \delta p=0}^{\text{standard}} \ne 0
\; \to \infty \;.
\eeq
This divergence can only be cancelled by
the angular counterterms, so that in this scheme they necessarily have to be
gauge dependent. Renormalizing $\alpha$ and $\beta$ through the
process-dependent scheme cancels all UV-divergent
gauge-dependent parts. The remaining UV-divergent gauge-independent
terms are then cancelled by $\delta m_{12}^2$. It can be defined
either via an $\overline{\mbox{MS}}$ condition or through the process
$H \to AA$. The overall NLO amplitude will finally be
gauge independent as it should be. \s

Applying the alternative tadpole scheme instead leads to the
cancellation of the UV-divergent gauge-dependent parts within the
residual amplitude, {\it i.e.}
\beq
\underline{\mbox {\it Alternative tadpole scheme:}} \qquad 
{\cal M}_{H \to hh}|_{\text{NLO}, \xi, \delta p=0}^{\text{tad}}
= 0 \;.
\eeq
The angular counterterms in turn can then be
defined gauge-independently. The unambiguous gauge-independent
definition of the angular counterterms is achieved through the pinched
scheme or the definition via a physical process. The counterterm for
$m_{12}^2$ is gauge-independent irrespective of the tadpole scheme and
can be renormalized in the $\overline{\mbox{MS}}$ or the
process-dependent scheme. \s 

We can summarize that a gauge-independent decay amplitude\footnote{We
  remind the reader that the schemes 
  previously proposed in the literature, relying on the application of
the standard tadpole scheme and a definition of the angular
counterterms through off-diagonal wave function renormalization
constants, lead to a manifestly gauge-dependent decay amplitude.} for the
process $H \to hh$ is achieved by applying the following
renormalization schemes for the angular counterterms:
\begin{center}\begin{tabular}{l|l|l}
tadpole treatment & $\delta \alpha, \delta\beta$ & 
gauge dependence $\delta \alpha$, $\delta \beta$ \\ \hline \hline
standard tadpole scheme & process dependent &
gauge dependent \\[0.2cm]
alternative tadpole scheme & pinched scheme &
gauge independent \\
& process dependent 
\end{tabular}\end{center}

Throughout the calculation we employ the alternative tadpole
scheme. This guarantees the manifestly gauge-independent
renormalization of the counterterms. It is furthermore indispensable
for a gauge-independent decay amplitude if the angular counterterms are not
obtained via a physical process. \s

Concerning a scheme with process-dependent counterterm definitions, note,
that the results for the NLO decay widths are
the same in the standard and in the alternative tadpole scheme. A
change of the tadpole scheme leaves the total NLO amplitude invariant,
it only moves around the gauge dependencies between the various building
blocks, so that in the alternative tadpole scheme the counterterms
become gauge independent.

\section{Numerical analysis \label{sec:numerics}}
The NLO EW corrections to the Higgs decay width $H\to hh$ have been
performed in two independent calculations and all results have been
cross-checked against each other. They agree within numerical
errors. The two computations use the {\tt
  Mathematica} package {\tt FeynArts 3.9} and {\tt 3.7}
\cite{Kublbeck:1990xc,Hahn:2000kx}, respectively, for the generation
of the LO and NLO amplitudes in the general $R_\xi$ 
gauge. For this, the model file for the CP-conserving 2HDM was used, which is
already implemented in the package\footnote{Note that the
  parametrization of the 2HDM potential implemented in the {\tt
    FeynArts} model file is different from the one presented in
  Section \ref{sec:modeldesc}. In particular instead of using
  $m_{12}^2$ the parameter $\Lambda_5 \equiv 2m_{12}^2 /(v^2 s_\beta
  c_\beta)$ is used. This has to be kept in mind when implementing the
  counterterm for $m_{12}^2$. }. The additionally needed tadpole and
self-energy amplitudes for the definition of the counterterms
and wave function renormalization constants have been generated in
the general $R_\xi$ gauge. For the contraction of the Dirac matrices
and the expression of the results in terms of scalar loop integrals
{\tt FeynCalc 8.2.0} \cite{Mertig:1990an,Shtabovenko:2016sxi} has been
applied in one calculation and {\tt FormCalc 8.1} \cite{Hahn:1998yk}
in the other. The {\tt C++} library {\tt LoopTools 2.12} and {\tt 2.9}
\cite{Hahn:1998yk}, respectively, has been used for the numerical
evaluation of the dimensionally regularized
\cite{'tHooft:1972fi,Bollini:1972ui} integrals. \s 

Our numerical evaluation has been performed with the following input
parameters. The fine structure constant $\alpha$ is taken at the
$Z$ boson mass scale \cite{Agashe:2014kda}, 
\beq
\alpha (M_Z^2) = \frac{1}{128.962} \;,
\eeq
and for the massive gauge boson masses we use
\cite{Agashe:2014kda,Denner:2047636} 
\beq
M_W = 80.385 \mbox{ GeV} \qquad \mbox{and} \qquad M_Z = 91.1876 \mbox{
GeV} \;.
\eeq
The lepton masses are chosen as \cite{Agashe:2014kda,Denner:2047636}
\beq
m_e = 0.510998928 \mbox{ MeV} \;, \quad
m_\mu = 105.6583715 \mbox{ MeV} \;, \quad
m_\tau = 1.77682 \mbox{ GeV} \;,
\eeq
and the light quark masses, following \cite{LHCHXSWG}, are set to
\beq
m_u = 100 \mbox{ MeV} \;, \quad m_d = 100 \mbox{ MeV} \;, \quad
m_s = 100 \mbox{ MeV} \;.
\eeq
The leptons and light quarks have only a small influence on the
results. For consistency with the ATLAS and CMS analyses the following
OS value for the top quark mass is taken,
\beq
m_t = 172.5 \mbox{ GeV} \;,
\eeq
as recommended by the LHC Higgs Cross Section Working Group
(HXSWG) \cite{Denner:2047636,Dittmaier:2011ti}. For the charm and
bottom quark OS masses we use \cite{Denner:2047636}
\beq
m_c = 1.51 \mbox{ GeV} \qquad \mbox{and} \qquad
m_b = 4.92 \mbox{ GeV} \;.
\eeq
As we do not include CP violation the CKM matrix is real, with the
CKM matrix elements given by \cite{Agashe:2014kda}
\beq
V_{\text{CKM}} = \left( \begin{array}{ccc} V_{ud} & V_{us} & V_{ub} \\
V_{cd} & V_{cs} & V_{cb} \\ V_{td} & V_{ts} & V_{tb} \end{array}
\right) = \left( \begin{array}{ccc} 0.97427 & 0.22536 & 0.00355 \\
    -0.22522 & 0.97343 & 0.0414 \\ 0.00886 & -0.0405 &
    0.99914 \end{array} \right) \;.
\eeq
Finally for the SM-like Higgs mass value, denoted by
$m_{H^\text{SM}}$, we take the most recent combined value from ATLAS
and CMS \cite{Aad:2015zhl}, 
\beq
m_{H^\text{SM}} = 125.09 \mbox{ GeV} \;. 
\eeq
In the 2HDM both the heavier and the lighter of the two CP-even Higgs
bosons can play the role of the SM-like Higgs boson, depending on the
chosen parameter set. In our investigated cases it is the
lighter of the CP-even Higgs bosons, $h$, that corresponds to $H^{\text{SM}}$. \s

For the numerical analysis only those 2HDM parameter sets have been
taken into account that have not yet been excluded by experimental and
the most relevant theoretical constraints. These parameter points have been
obtained by scans performed in the 2HDM parameter space with the tool
{\tt ScannerS} \cite{Coimbra:2013qq}.\footnote{We are indebted to Marco
  Sampaio, one of the authors of {\tt ScannerS}, for generously
  providing us with valid parameter sets.} It checks if the chosen
CP-conserving vacuum represents the global minimum
\cite{Barroso:2013awa}, if the 2HDM potential is bounded from
below\cite{Deshpande:1977rw} and if tree-level unitarity
holds \cite{Kanemura:1993hm,Akeroyd:2000wc}. The consistency with the
electroweak precision constraints
\cite{Peskin:1991sw,PhysRevD.45.2471,Grimus:2008nb,Haber:2010bw,ALEPH:2010aa,Baak:2011ze,Baak:2012kk}
is assumed to be fulfilled if the $S,T$ and $U$ variables
\cite{Peskin:1991sw} predicted by the 2HDM are within the 95\% ellipsoid
centered on the best fit point to the EW data. Loop processes with
charged Higgs bosons induce indirect constraints that depend on $\tan\beta$
via the charged Higgs coupling to the fermions. They dominantly stem
from $B$ physics observables
\cite{Mahmoudi:2009zx,Deschamps:2009rh,Hermann:2012fc} and the
measurement of $R_b$
\cite{Denner:1991ie,Grant:1994ak,Haber:1999zh,Freitas:2012sy}. 
In our analysis we take the most recent bound of $m_{H^\pm} \gsim 480$~GeV for the 
type II and flipped 2HDM \cite{Misiak:2015xwa}. Note, that the results
from LEP \cite{Abbiendi:2013hk} and the LHC
\cite{Aad:2014kga,Khachatryan:2015qxa}\footnote{The recent ATLAS
  results \cite{Aad:2015typ}  have not been translated into 
  bounds so far.} require the charged Higgs mass to be above ${\cal O}
(100 \mbox{ GeV})$ depending on the model type. For the check of the
compatibility with the LHC Higgs data {\tt 
  ScannerS} uses the Higgs production cross sections through gluon fusion and
$b$-quark fusion at NNLO QCD, which are obtained from an interface
with {\tt SusHi} \cite{Harlander:2012pb}. The remaining production
cross sections are taken at NLO QCD \cite{LHCHXSWG}, and the 2HDM
Higgs decays are computed with {\tt HDECAY}
\cite{Djouadi:1997yw,Harlander:2013qxa}. The EW corrections are
consistently neglected in the computation of these processes as they have not been
provided for the 2HDM so far. The program  {\tt HiggsBounds}
\cite{Bechtle:2008jh,Bechtle:2011sb,Bechtle:2013wla} is used for the
check of the exclusion limits and {\tt HiggsSignals}
\cite{Bechtle:2013xfa} is used to test the compatibility with the
observed signal for the 125~GeV Higgs. Further details can be found in
\cite{Ferreira:2014dya}. All results shown in the following analysis are for
the 2HDM type II. \s

For the numerical analysis we exploit three different sets of
parameter points that are distinguished with respect to their Higgs
spectra but that all fulfill the above listed experimental and theoretical
constraints:
\begin{itemize}
\item[(i)] The parameter sets are chosen such that the decay $H\to hh$
  is kinematically possible, hence
\beq
{\mbox{\it Condition (i): }} \quad M_H \stackrel{!}{\ge} 2 M_h \;.
\eeq
\item[(ii)] The parameter sets are chosen such that the decay $H\to hh$
  is kinematically possible. Additionally, we require the heavy Higgs boson masses to
  maximally deviate by $\pm 5\%$ from $M$, with $M^2 \equiv m_{12}^2/(s_\beta
  c_\beta)$. We hence have 
\beq
{\mbox{\it Condition (ii): }} && M_H \stackrel{!}{\ge} 2 M_h \qquad\qquad \mbox{
and } \\
&&  m_{\phi_{\text{heavy}}} \stackrel{!}{=} M \pm 5\% \; , \; \mbox{ with }
m_{\phi_{\text{heavy}}} \in \{ m_H, m_A, m_{H^\pm}
\} \label{eq:mpm5pc} 
\;.
\eeq
In these scenarios the non-SM Higgs bosons are approximately mass
degenerate and of the order of the ${\mathbb Z}_2$ breaking scale. 
\item[(iii)] The conditions for the parameter sets chosen here are
  that both the decay $H\to hh$ and the decay $H\to AA$ are kinematically
  possible, {\it i.e.}
 \beq
{\mbox{\it Condition (iii): }} \quad M_H \stackrel{!}{\ge} 2 M_h
\qquad \mbox{and} \qquad M_H \stackrel{!}{\ge} 2 M_A \;.
\label{eq:cond3}
\eeq
\end{itemize}

As we have seen in subsection \ref{subsec:Htohh} the decay $H \to hh$
depends through the Higgs self-coupling $\lambda_{Hhh}$ on both mixing
angles $\alpha$ and $\beta$ and on the soft $\mathbb{Z}_2$ breaking parameter
$m_{12}^2$. This process hence allows us to study the numerical
stability of the renormalization schemes for the mixing angles but in
particular also of the mass parameter $m_{12}^2$. The possible
renormalization schemes for the angular 
counterterms are denoted as follows, 
\beq
\begin{array}{lll}
\mbox{proc} &:& \quad \mbox{process-dependent}
\\
p_\star^{c,o} &:& \quad p_\star \mbox{ tadpole-pinched},\; \delta
\beta^{(1)} \mbox{ ('$c$') or } \delta \beta^{(2)} \mbox{ ('$o$')} \\
\mbox{pOS}^{c,o} &:& \quad \mbox{on-shell tadpole-pinched},\;
\delta\beta^{(1)} \mbox{ ('$c$') or } \delta \beta^{(2)} \mbox{
  ('$o$')} \;.
\end{array}
\label{eq:renschemenotation}
\eeq
As explained above, the process-dependent renormalization for $\alpha$
proceeds through the decay $H \to \tau\tau$ and the one for $\beta$
exploits $A\to \tau\tau$. In the tadpole-pinched schemes,
$p_\star$ or pOS, $\beta$ can be
renormalized through the charged sector, with the counterterm denoted by $\delta
\beta^{(1)}$, or through the CP-odd sector, with the counterterm given
by $\delta \beta^{(2)}$. For $m_{12}^2$ we adopt the two schemes
\beq
\begin{array}{lll}
\mbox{proc} &:& \quad \mbox{process-dependent via } H\to AA \mbox{ and}
\\
\overline{\mbox{MS}} &:& \quad \mbox{modified minimal subtraction
  scheme} \;.
\end{array}
\eeq
We investigate the size of the NLO corrections by defining
\beq
\Delta \Gamma \equiv \frac{\Gamma^{\text{NLO}}-
  \Gamma^{\text{LO}}}{\Gamma^{\text{LO}}} \;.
\eeq
This ratio measures the relative size of the NLO corrections compared
to the LO decay width. We start by investigating the impact of the
angular renormalization schemes on the NLO corrections to the
Higgs-to-Higgs decays. To this end we
show in Fig.~\ref{fig:angrenschemes} for all parameter sets of $(i)$
the relative NLO corrections $\Delta \Gamma^{H\to hh}$ as a function
of the LO width for all 
possible angular schemes defined in (\ref{eq:renschemenotation}). 
For $\beta$ both possible renormalization choices through the charged
and through the CP-odd sector have been applied in the tadpole-pinched
schemes. For $\delta m_{12}^2$ the $\overline{\mbox{MS}}$ scheme has
been applied with the renormalization scale set to $\mu_R = 2 m_h$. 
As can be inferred from the plot, the relative corrections can be 
huge. Discarding the region for small LO widths, where $\Delta
\Gamma^{H\to hh}$ diverges\footnote{While the NLO
    width also tends to 
  zero when the LO width becomes small, for some parameter
  configurations there remains a non-zero NLO width also for 
  $\Gamma^{\text{LO}}=0$, due to cancellations among various terms
  contributing at NLO.}, we have relative 
corrections of up to about 400\% (not shown in the plot) 
in the process dependent scheme and
of up to about 200\% for the tadpole-pinched
schemes. Note that we cut the plot at $\Delta
  \Gamma^{Hhh}= -100\%$ in order to avoid negative widths. \s

The appearance of huge corrections is not necessarily due to
numerical instability. It is rather the non-decoupling effects,
generically arising in the 2HDM 
\cite{Kanemura:2004mg,Kanemura:2002vm},
that blow up the NLO corrections. This shall be explained in the
following. For the decay $H\to hh$ being kinematically possible large
enough $H$ masses are needed. As can be read off from
Eq.~(\ref{eq:massrels}), heavy masses can either be obtained through a large mass
parameter $M$ or through the VEVs. They enter the mass relation with a coefficient
proportional to a linear combination of the Higgs potential couplings
$\lambda_i$. In the {\it decoupling limit} we have $c_\phi^2 M^2 \gg
f(\lambda_i) v^2$, and the spectrum effectively consists of heavy Higgs
bosons whose masses are given by the scale $c_{\phi_{\text{heavy}}}
M$ independently of the $\lambda_i$, and of 
one light resonance that represents the SM-like Higgs boson. 
The trilinear and quartic scalar couplings controlled by  $\lambda_i$
are comparatively small  
and all loop effects due to the heavier Higgs bosons vanish in the limit
$m_{\phi_{\text{heavy}}}^2 \to \infty$ because of the decoupling theorem
\cite{Appelquist:1974tg}. This situation corresponds to the decoupling limit of the
MSSM, where supersymmetry requires the couplings $\lambda_i$ to be
replaced by the gauge couplings $g$ and $g'$ and where heavy masses can only
be obtained through a large mass scale $M$ usually chosen to be the
pseudoscalar mass $M_A$.
In the opposite case, the {\it strong coupling
regime}, we have $c_{\phi_{\text{heavy}}}^2 M^2 \lsim f(\lambda_i) \,
v^2$ for at least one of the non-SM-like Higgs bosons, and large mass values 
can only be obtained for large couplings $\lambda_i$. The decoupling
theorem does not apply and the radiative corrections of the heavy
Higgs bosons develop a power-like behaviour in
$m_{\phi_{\text{heavy}}}$, also known as
non-decoupling effects  
\cite{Ciafaloni:1996ur,Kanemura:1997wx,Ginzburg:1999fb,Kanemura:1999tg,Arhrib:1999rg,Kanemura:1997ej,Ginzburg:2001ph,Malinsky:2002mq,Arhrib:2003ph,Malinsky:2003bd}. They
grow proportional to $m_{\phi_{\text{heavy}}}^4$
\cite{Kanemura:2004mg,Kanemura:2002vm}. The huge corrections in
Fig.~\ref{fig:angrenschemes} are due to this power law for scenarios
with heavy non-SM Higgs bosons. \s
\begin{figure}[t!]
\begin{center}
\includegraphics[width=250pt, trim = 0mm 0mm 0mm 0mm, clip]{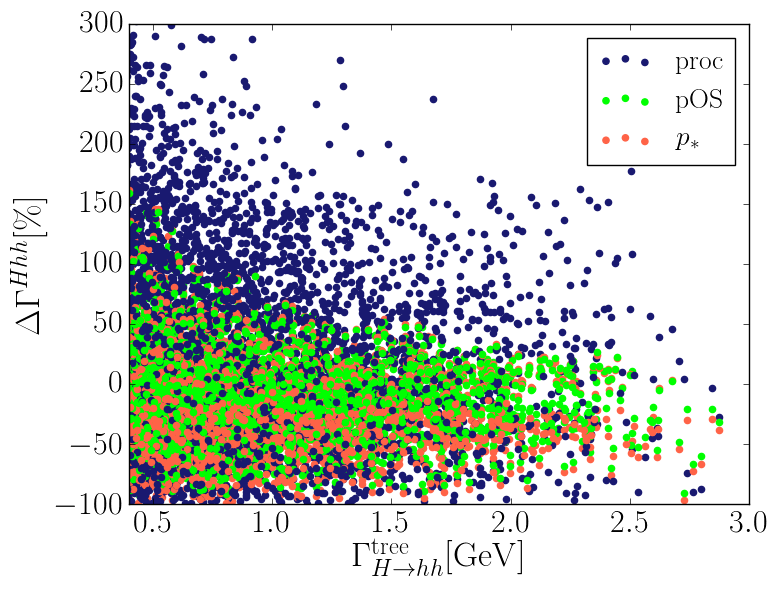}
    \caption{Scatter plot for the relative NLO corrections to $H\to
      hh$ for all parameter points passing the theoretical and
      experimental constraints and fulfilling the kinematic condition
      $(i)$, as a function of the LO width; shown for various
      angular renormalization schemes: process-dependent (blue), pOS
      tadpole-pinched (green), $p_\star$ tadpole-pinched (red);
      $m_{12}^2$ has been $\overline{\mbox{MS}}$ renormalized with
      $\mu_R=2 m_h$. Scenarios leading to negative widths for one of
      the renormalization schemes have been discarded, and we have cut
      at 300\% for positive corrections.}
\label{fig:angrenschemes}
\end{center}
\end{figure} 

From the above discussion it becomes clear that 
for a meaningful discussion of the numerical
stability of the different renormalization schemes we have to separate
the two effects: huge corrections due to large
couplings $\lambda_i$ and corrections that are blown-up due to numerical
instability of the chosen renormalization scheme. 
We therefore investigate the relative NLO
corrections for the parameter set $(ii)$ where we require all non-SM
heavy Higgs masses to lie within 5\% around the mass scale $M$ set by
the soft $\mathbb{Z}_2$ breaking mass parameter. In this limit, the loop effects
of the heavy particles are expected to decouple. However, even if  
Eq.~(\ref{eq:mpm5pc}) is fulfilled, the decoupling does not
necessarily take place. It is found to be impossible, in fact, in the
limit $s_{\alpha+\beta}  \to 1$. This limit is referred to as the {\it
wrong sign limit} as for the 2HDM type II (and F) it implies a relative
minus sign in the couplings of the SM-like Higgs boson to down-type fermions
with respect to its couplings to massive 
gauge bosons (and up-type fermions) \cite{Ferreira:2014dya,Ferreira:2014naa,Ferreira:2014sld,Ferreira:2014qda}. In Ref.~\cite{Ferreira:2014naa} it was shown
that non-decoupling properties inevitably arise for $s_{\beta+\alpha}
\to 1$ in the 2HDM. The non-decoupling of charged Higgs contributions
in the loop induced $h\gamma \gamma$ coupling was also discussed in
\cite{Dumont:2014wha,Fontes:2014tga,Fontes:2014xva}. \s 

In order to examine the non-decoupling
properties of the loop contributions to $H \to hh$ we focus on the
trigonometric relations relevant for the involved Higgs couplings. Two
limiting cases are of interest, given by $s_{\beta-\alpha} \approx 1$
and $s_{\beta -\alpha} <1$. While $s_{\beta-\alpha} \to 1$ corresponds
to the SM limit, in the wrong sign regime significant deviations from
this limit are still compatible with LHC data. Thus it was shown in
\cite{Ferreira:2014dya,Ferreira:2014sld,Ferreira:2014qda} that values
of $s_{\beta-\alpha} \approx 0.55$ (0.62) are 
compatible with the LHC Higgs data at 3 (2)$\sigma$ and additionally
fulfill the other constraints tested by {\tt ScannerS}. Relatively
small values of $s_{\beta-\alpha}$, however, require significant
contributions from the second term in Eq.~(\ref{eq:massrels}), given by
$f(\lambda_i) v^2$, even if $m_H^2 \approx M^2$, in order to acquire a
sufficiently large $m_H$ for the decay $H \to hh$ to take place. This,
however, drives us back to the non-decoupling limit. \s

Also in the limit $s_{\beta-\alpha} \to 1$,
corresponding to SM-like $h$  
couplings to the massive gauge bosons, the trilinear coupling
$\lambda_{Hhh}$ can become large in the wrong sign limit. Analogous
to the non-decoupling of the charged Higgs contribution in the decay
$h \to \gamma \gamma$ studied in Ref.~\cite{Ferreira:2014naa}, also
the other heavy 
Higgs bosons $H$ and $A$ exhibit a non-decoupling behaviour in the
wrong sign limit. In order to show this, we consider the ratio
$\lambda_{HHh}/m_H^2$, which plays a role in the EW corrections to $H\to
hh$. We analyze this ratio for both the correct and the wrong sign
regime in the limit $s_{\beta-\alpha}\to 1$, where $m_H^2 \approx
M^2$. In the wrong sign regime, where 
  $s_{\beta+\alpha} \to 1$, $t_\beta$ has to be large in order to come
  close to the SM limit. We thus obtain
\beq
\frac{\lambda_{HHh}}{m_H^2} &=& - \frac{1}{m_H^2 v}
\frac{s_{\beta-\alpha}}{s_{2\beta}} \left[ s_{2\alpha} (2 m_H^2 +
  m_h^2) - M^2 (3s_{2\alpha} + s_{2\beta} ) \right] \nonumber \\
&\approx&
\frac{1}{v} \frac{s_{\beta - \alpha}}{s_{2\beta}} [ s_{2\alpha} +
s_{2\beta} ] + {\cal O} \left( \frac{m_h^2}{vm_H^2}\right) \nonumber
\\
&=& \frac{1}{v} s_{\beta-\alpha} \left( 1 - \frac{s_{\beta-\alpha} -
    c_{\beta-\alpha} t_\beta}{s_{\beta-\alpha} + c_{\beta-\alpha}
    t_\alpha} \right) + {\cal O} \left( \frac{m_h^2}{v m_H^2} \right)
\nonumber \\[0.1cm]
&& \hspace*{-0.7cm} \left\{  \begin{array}{cll} 
\stackrel{s_{\beta-\alpha} \to 1}{\approx} & 0 & \;\mbox{ correct sign
  limit} \\ 
\stackrel{{\scriptsize \begin{array}{l}s_{\beta+\alpha} \to 1, \\ t_\beta \to
    \infty\end{array}}}{\approx} &
2/v & \;\mbox{ wrong sign limit} 
\end{array} \right. + {\cal O} \left( \frac{m_h^2}{v m_H^2} \right)
\;.
\label{eq:limit}
\eeq
As can be inferred from Eq.~(\ref{eq:limit}) the ratio
$\lambda_{HHh}/m_H^2$ approaches a constant value in the wrong sign
regime so that the heavy Higgs loop contributions do not decouple for
$m_H^2 \to \infty$. In contrast, in the correct sign limit the ratio
vanishes and the decoupling of heavy $H$ loop effects takes
place. Analogously, the ratio $\lambda_{Hhh}/m_H^2$ yields a constant
value in the wrong sign regime and prevents a decoupling of heavy
loop particle contributions. The same holds for $\lambda_{hAA}/m_A^2$
where we find
\beq
\frac{\lambda_{hAA}}{m_A^2} &=& \frac{2}{v} c_{\beta-\alpha} \left(
  \frac{1}{t_\beta} - t_\beta \right) + {\cal O} \left( \frac{m_h^2}{v
  m_A^2} \right) \nonumber \\
\nonumber \\[0.1cm]
&& \hspace*{-0.7cm} \left\{  \begin{array}{cll} 
\stackrel{s_{\beta-\alpha} \to 1}{\approx} & 0 & \;\mbox{ correct sign
  limit} \\ 
\stackrel{{\scriptsize \begin{array}{l}s_{\beta+\alpha} \to 1, \\ t_\beta \to
    \infty\end{array}}}{\approx} &
-2/v & \;\mbox{ wrong sign limit} 
\end{array} \right. + {\cal O} \left( \frac{m_h^2}{v m_A^2} \right)
\;.
\eeq
This non-decoupling behaviour in the wrong sign regime explains why
even in the case where the heavy Higgs boson masses are controlled by
\begin{figure}[t!]
  	\centering
\includegraphics[width=225pt,clip]{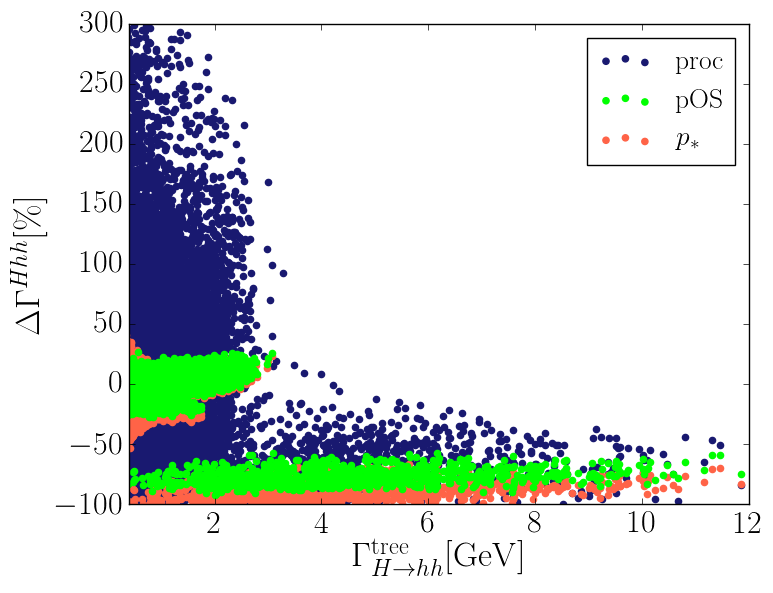}
\includegraphics[width=225pt,clip]{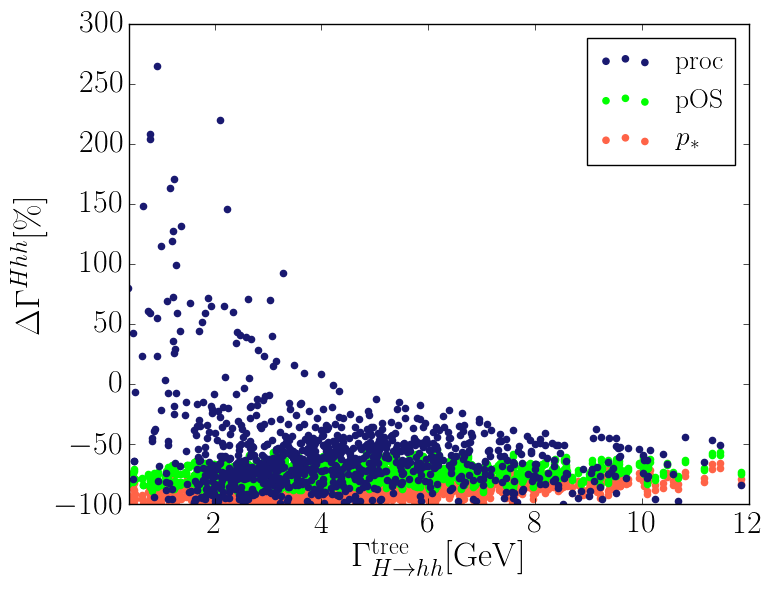}
    \caption{Scatter plot for the relative NLO corrections to $H\to
      hh$ for all parameter points passing the theoretical and
      experimental constraints and fulfilling the kinematic condition
      $(ii)$, as a function of the LO width; shown for various
      angular renormalization schemes: process-dependent (blue), pOS
      tadpole-pinched (green), $p_\star$ tadpole-pinched (red);
      $m_{12}^2$ has been $\overline{\mbox{MS}}$ renormalized with
      $\mu_R = 2m_h$. Scenarios leading to
        negative widths for one of the renormalization schemes have
        been discarded, and we have cut at 300\% for positive
        corrections. Left: all points, right: only those with 
        $\sin\alpha >0$, corresponding to the wrong sign regime.}
\label{fig:scen2all}
\end{figure} 
the mass parameter $M$ the loop effects do not decouple and give
rise to large radiative corrections. This behaviour is shown in the
following plots. In Fig.~\ref{fig:scen2all} (left) we first
  display for {\it all} points of parameter set $(ii)$ that pass the
  theoretical and experimental constraints the relative NLO
  corrections as a function of the LO width for the process-dependent
  and the two tadpole-pinched schemes in the angular
  renormalization. For $m_{12}^2$ $\overline{\text{MS}}$
  renormalization has been applied at $\mu_R = 2 m_h$. Although our
  involved heavy Higgs masses are due to a large value of $M$, we
  observe huge relative corrections of up to 300\% and larger. Note that in the
  plot we cut at -100\% in order to avoid negative widths. Following
  our considerations on the decoupling behaviour of loop corrections
  in the SM-limit, we now divide our parameter points into those
  of the wrong sign regime, where $s_{\beta+\alpha}
    \approx1$, and those 
  of the correct sign regime with $s_{\beta-\alpha} \approx 1$. 
  We used the sign of $s_{\alpha}$ as discriminator between the two regimes, collecting the parameter sets with $s_\alpha > 0$ for the former
and the ones with $s_\alpha < 0$ for the latter case.\footnote{We explicitly
    verified that for $\sin\alpha <0$ the ratio of involved coupling
    over corresponding loop mass is relatively small, while the sets with $\sin
    \alpha >0$ comprise ratios with much larger values, reflecting the
  non-decoupling situation.} 
This leads to
  Fig.~\ref{fig:scen2all} (right) which displays the relative
  corrections for the wrong sign regime and
  Fig.~\ref{fig:scen2corrsign} for the correct sign regime. We show
  results for all applied renormalization schemes, 
  discard points with small LO widths and cut at +300\% and
  -100\%, the latter to avoid negative widths. 
As expected, in
  Fig.~\ref{fig:scen2all} (right), despite the fact that all heavy 2HDM Higgs masses 
have been chosen within 5\% around $M$, the corrections can be huge,
reaching up to 300\% and larger (not shown in the plot). The  
plot shows that in the tadpole pinched schemes for the
displayed parameter points\footnote{Including also
  scenarios with relative corrections beyond 100\%, the relative corrections in
  the tadpole pinched schemes can also be larger.} the relative corrections for all
scenarios are within about -50\% to -100\%. In the
process-dependent 
scheme we can have rather small corrections, but also huge
corrections, exceeding largely those of the process-independent
schemes. Large corrections as found for the tadpole pinched schemes
are to be expected for significant coupling strengths as involved in 
the NLO diagrams here. This is confirmed by the explicit verification that in this
non-decoupling regime the pure vertex corrections become large. The
small corrections found for some scenarios in the process-dependent
scheme are due to accidental cancellations between the various terms
contributing at NLO and not because of more numerical stability in
this renormalization scheme. This is why we observe here also huge
corrections of up to 300\% and beyond while this is not the case for the
process-independent schemes. In order to be able to draw more
conclusive statements on the numerical stability, corrections beyond
the one-loop level would have to be calculated in this regime of
strong coupling constants. This is beyond the scope of this paper. \s

\begin{figure}[t!]
\centering
\includegraphics[width=225pt, trim = 0mm 0mm 0mm 0mm,
        clip]{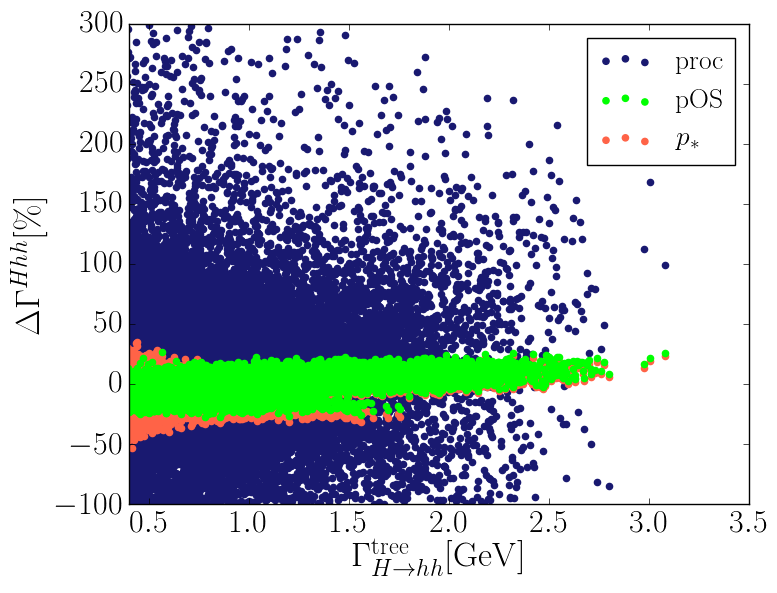}
\includegraphics[width=225pt, trim = 0mm 0mm 0mm
0mm,clip]{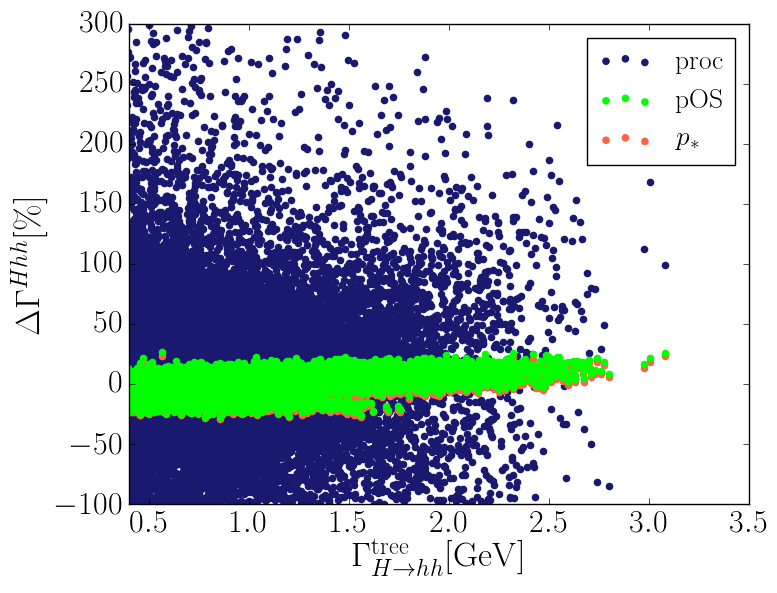}
\caption{Same as Fig.~\ref{fig:scen2all} but only with points featuring
      scenarios in the correct sign limit, {\it
          i.e.}~$\sin \alpha <0$. Left: all schemes, right: without
        $\beta$ renormalization in the $p_\star^o$ scheme, see text.} 
\label{fig:scen2corrsign}
\end{figure} 
Taking into account only scenarios in the correct sign limit, we are left with
Fig.~\ref{fig:scen2corrsign}, where we cut on scenarios leading to
relative corrections beyond $+ 300$\% and $-100$\%, respectively, and
discarded those with small LO widths. As explained above in detail, we are now
truly in the decoupling limit. This is reflected by the plot. Since the
involved trilinear couplings are not as large as in the wrong sign
regime, for the process-independent renormalization schemes the
relative NLO corrections have become considerably smaller as compared
to the wrong sign case. Also of course, the LO widths are
smaller.\footnote{Note the different scales in
    Fig.~\ref{fig:scen2all} and Fig.~\ref{fig:scen2corrsign}.} Having
excluded scenarios with enhanced corrections due to non-decoupling, we can now
proceed to investigate the numerical stability of the applied
schemes. Inspecting Fig.~\ref{fig:scen2corrsign} (left), we see that
the corrections in the tadpole-pinched schemes all lie between 
about -60 and +40\%. The process-dependent
renormalization on the other hand induces much larger corrections, of
the order of up to 300\% and larger. While again we can also have
small corrections in the 
process-dependent scheme, this is due to accidental cancellations and
not a sign of numerical stability. This statement is underlined by the fact that
the corrections in this scheme can become huge as well, whereas in the
process-independent schemes they do not exceed $-60$\%.
In Fig.~\ref{fig:scen2corrsign} (left) we furthermore see a difference
between the pOS and the 
$p_\star$ tadpole-pinched scheme. 
For small LO widths the relative NLO corrections in the $p_\star$ tadpole-pinched
scheme increase more quickly. This behaviour can be traced back to
the appearance of the top resonance in the $G^0 A$ self-energy encountered
in the $\beta$ renormalization through the CP-odd Higgs sector, {\it
  i.e.}~in $\delta \beta^{(2)}$. For masses $m_A^2/2 = 4 m_t^2$ the
diagram shown in Fig.~\ref{fig:selfendiag} becomes resonant. 
This requires relatively light pseudoscalar masses of about 488 GeV. The
tail of this effect is, however, still visible for masses up to 
$m_A \approx 700$~GeV. 
In the renormalization through the charged sector no self-energy diagrams
with pure top loop contributions are encountered in the mixed $G^\pm
H^\pm$ self-energy so that the counterterm $\delta \beta^{(1)}$ is not
affected by the top resonance. Note furthermore that the counterterm
$\delta \beta^{(2)}$ in the pOS scheme would require $A$ masses as low
as 350~GeV to hit the top resonance. These are not included in set
$(ii)$ so that no resonant enhancement is visible in the pOS
scheme. \s

\begin{figure}[t!]
\vspace*{-0.5cm}
  	\centering
\includegraphics[width=150pt, trim = 0mm 5mm 0mm 4mm,
clip]{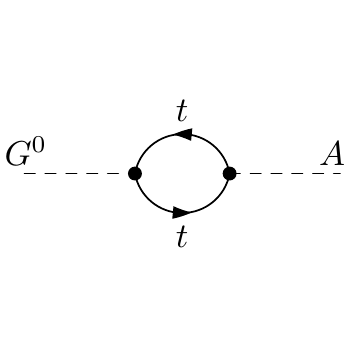}
\vspace*{-0.5cm}
    \caption{Top loop diagram contributing to the mixed self-energy 
    $\Sigma^{\text{tad}}_{G^0 A}$ in the $\beta$
    renormalization.}
\label{fig:selfendiag}
\end{figure} 
In Fig.~\ref{fig:scen2corrsign} (right) we have excluded the $p_\star^o$
renormalization of $\delta \beta$ from the plot. As expected all
tadpole-pinched schemes now show the same behaviour. 
For scenarios with light pseudoscalar masses the $\beta$
renormalization through the charged sector might therefore be preferable. 
From these investigations we furthermore conclude that the
tadpole-pinched schemes are numerically stable and can hence be 
advocated as renormalization schemes for the mixing angles that are
numerically stable, gauge independent and process independent. This
confirms our findings of \cite{Krause:2016oke} in a process involving
a coupling that has a complicated dependence on $\alpha$ and $\beta$
so that the cancellation of huge tadpole contributions is
non-trivial. Moreover, the plots show the good numerical behaviour of
the $\overline{\mbox{MS}}$ scheme applied for $\delta m_{12}^2$. 
Independently of the discussion with respect to
  numerical stability we have seen that also in the tadpole-pinched
  schemes the corrections can be significant due to non-decoupling
  behaviour of the corrections. In these cases clearly higher order
  corrections have to be included in order to make reliable
  predictions. This is beyond the scope of this paper.
\s

We finalize the discussion of the angular counterterms by examining a
specific scenario in the decoupling
limit\footnote{The masses of $A$ 
  and $H^\pm$ do not deviate by more than 5\% from $M$. The heavy
  Higgs mass $m_H$ deviates by 5.7\% at the lowest and by
  20\% at the highest mass value in the chosen range.}. It is given by
\beq
\begin{array}{llll}
\hspace*{-0.2cm} \mbox{\it \underline{Scen1:}} & \; m_{H} =
(671.05...803.12) \mbox{ GeV}\,, & \; m_A = 700.13 \mbox{ GeV}\,, & \;
m_{H^\pm} =700.35 \mbox{ GeV}\,, \\[0.1cm] 
\hspace*{-0.2cm} & \; \tan\beta = 1.45851 \,, & \; \alpha = -0.570376 \,, & \;
m_{12}^2 = 2.0761\cdot 10^5 \mbox{ GeV}^2 
\label{eq:scen1}
\end{array}
\eeq
\begin{figure}[t!]
  	\centering
  	\includegraphics[width=250pt, trim = 0mm 0mm 0mm 0mm,
        clip]{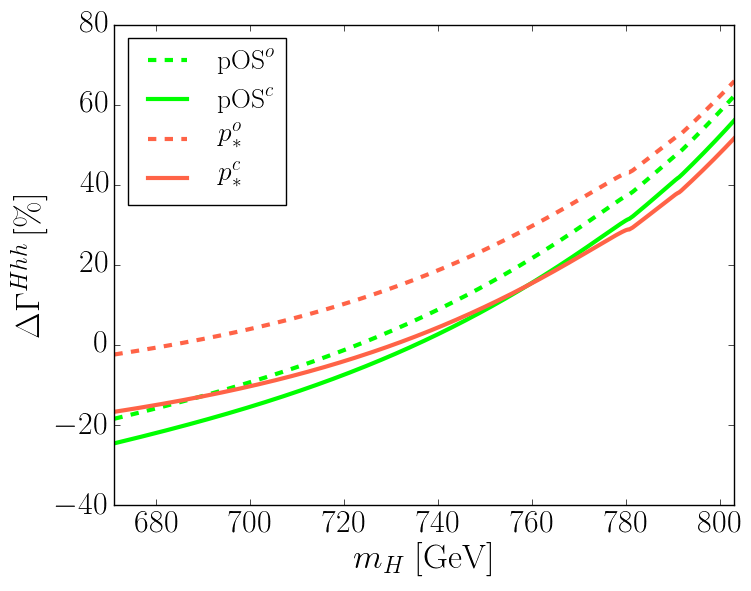}
    \caption{Relative NLO corrections to $H\to
      hh$ for angular renormalization in the tadpole-pinched schemes as
      defined in Eq.~(\ref{eq:renschemenotation}), with the 2HDM
      parameters given by {\it Scen1}, Eq.~(\ref{eq:scen1}). 
      $m_{12}^2$ has been $\overline{\mbox{MS}}$ renormalized with
      $\mu_R = 2m_h$.}
\label{fig:scen1}
\end{figure} 
The chosen pseudoscalar Higgs mass is far above the top resonance so
that no enhanced contributions in the $p_\star$ scheme are to be expected. Figure
\ref{fig:scen1} displays the relative NLO correction to the decay
$H\to hh$ for {\it Scen1} as a function of the heavy Higgs boson mass
$m_H$ for the renormalization of the mixing angles in the $p_\star$
and in the OS tadpole-pinched schemes. The angle $\beta$ has been
renormalized both through the charged and through the CP-odd sector. 
We do not include the numerically unstable process-dependent renormalization.
The kinks in the curves which appear independently of the renormalization scheme at
$m_H \approx 781$~GeV and 791 GeV (not visible in the plot) are due to
threshold effects in the scalar 
two-point function $B_0$ appearing in the counterterms. They are given
by the following parameter configurations and counterterms
\begin{center}
\begin{tabular}{ccccc}  
\toprule
Kink & Kinematic point & Origin \\
\midrule
1 & ~ $m_H (780.74 \mbox{ GeV}) = m_{H^\pm} (700.34 \mbox{ GeV}) +
M_W$~ & $\delta Z_{hh}, \delta Z_{HH}, \delta Z_{Hh}$ \\[0.1cm]
2 & ~ $m_{H} (791.31 \mbox{ GeV}) = m_A (700.13 \mbox{ GeV}) + M_Z$~ &
$\delta Z_{hh}, \delta Z_{HH}, \delta Z_{Hh}$ \\ 
\bottomrule
\end{tabular}
\end{center} 
In the investigated mass range the LO width varies between 0.356~GeV
at the lowest and 0.221~GeV at the highest $m_H$ value. 
As can be inferred from the plot, the relative corrections range
between about -25\% and +66\% depending on $m_H$ and on the
renormalization scheme. The corrections are
large, but not numerically unstable. Comparing the results in
the $p_\star^c$ and $p_\star^o$ scheme and those of the pOS$^c$ and
pOS$^o$ scheme, the remaining theoretical uncertainty due to missing
higher order corrections can be estimated based on a change of the
renormalization scheme for $\beta$. The $p_\star$ scheme is more
affected by the change of the renormalization scheme and induces an
estimated theoretical uncertainty which varies between about 17\% and
9\%  from the lower to the upper $m_H$ range. The residual theoretical
uncertainty can also be estimated from the scale change by comparing
the pOS$^o$ with the $p_\star^o$ scheme on the one hand and
pOS$^c$ with the $p_\star^c$ scheme on the other hand. In the lower
mass range the $\beta$ renormalization through the CP-odd sector
suffers more from a change of the renormalization scale than the one
through the charged Higgs sector. For the former the theoretical
uncertainty is estimated to be about 20\% here. At $m_H=803$~GeV for both
schemes the uncertainties are similar with about 2-3\%. Note that with
growing $m_H$ the scenario departs more and more from the decoupling
regime which is reflected in the increase of the NLO corrections. \s

So far we have used the renormalization scale $\mu_R=2m_h$ in the
$\overline{\text{MS}}$ renormalization of $\delta m_{12}^2$. This scale
choice is justified by Fig.~\ref{fig:renscale}. It shows the relative NLO
corrections for the parameter points of set $(i)$ with $m_{12}^2$
$\overline{\mbox{MS}}$ renormalized at three different renormalization
scales, given by $\mu_R = m_H$, $m_h$ and $2
m_h$. Scenarios with small LO widths have been
  discarded, and we have cut the relative negative and positive
  corrections at -100\% and 300\%, respectively.
The angles have been renormalized in the OS tadpole-pinched scheme. 
\begin{figure}[t!]
\centering
\includegraphics[width=250pt,
clip]{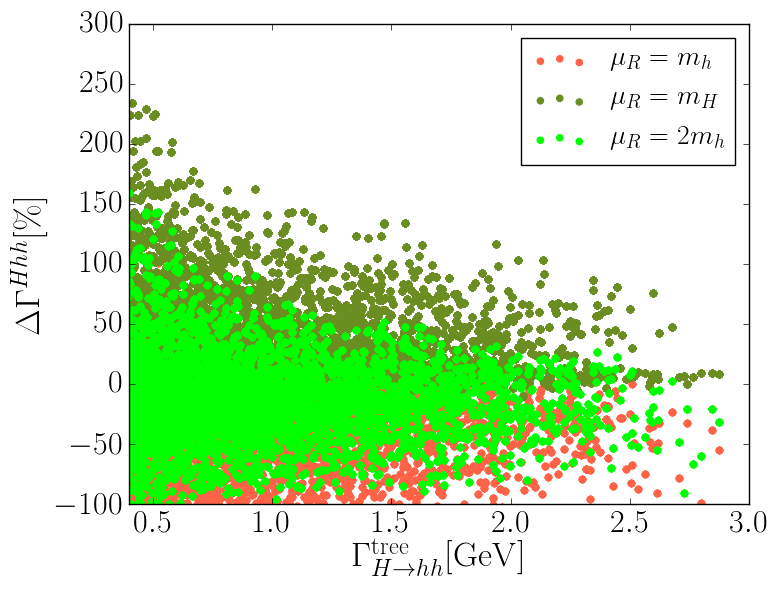}
   \caption{Scatter plot for the relative NLO corrections to $H\to
      hh$ for all parameter points passing the theoretical and
      experimental constraints and fulfilling the kinematic condition
      $(i)$, as a function of the LO width; shown for three different
      renormalization scales in the $\overline{\mbox{MS}}$
      renormalized $m_{12}^2$: $\mu_R = m_h$ (red), $m_H$ (dark green)
      and $2 m_h$ (light green); 
      angles are pOS tadpole-pinched
      renormalized. Scenarios with negative NLO
        widths have been excluded, and the relative positive
        corrections have been cut at 300\%.}
\label{fig:renscale}
\end{figure} 
As can be inferred from the plot, $\mu_R = 2m_h$ yields the smallest
corrections and is hence the recommended scale among the
three. \s

\begin{figure}[b!]
  	\centering
 \includegraphics[width=250pt, trim = 0mm 0mm 0mm 0mm,
        clip]{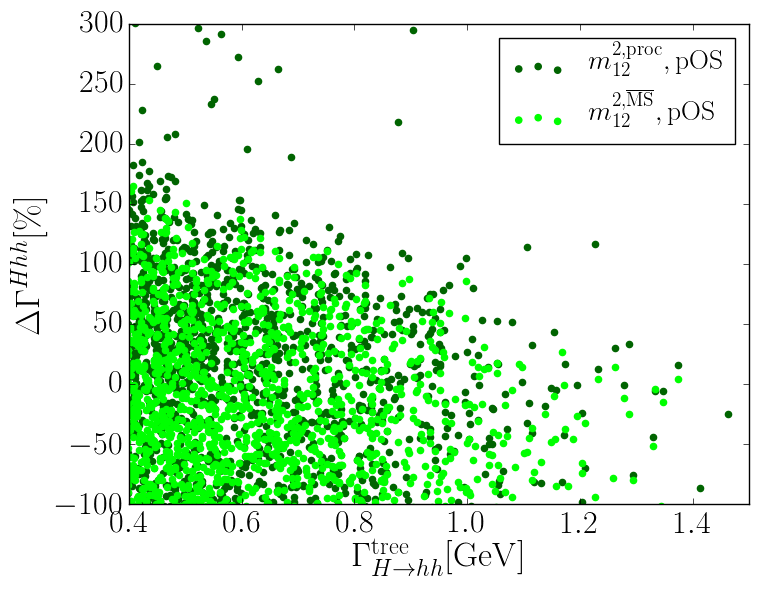}
    \caption{Scatter plot for the relative NLO corrections to $H\to
      hh$ for all parameter points passing the theoretical and
      experimental constraints and fulfilling the kinematic condition
      $(iii)$, as a function of the LO width; shown for the
      process-dependent renormalization of $m_{12}^2$ (dark green) and
      $\overline{\mbox{MS}}$ renormalization with $\mu_R = 2m_h$. The
      angles have been renormalized in the pOS
      scheme. Scenarios with negative NLO
        widths have been excluded, and the relative positive
        corrections have been cut at 300\%.} 
\label{fig:set3scatter}
\end{figure} 
We now turn to the investigation of the process-dependent
renormalization of $m_{12}^2$. For this purpose we use the parameter
points of set $(iii)$ for which $H \to AA$ decays are
kinematically allowed. Clearly, here we are not in the decoupling
regime any more due to the mass hierarchy among the heavy non-SM
Higgs bosons, so that large radiative corrections are to be
expected. This is confirmed by Fig.~\ref{fig:set3scatter} which shows
the relative NLO corrections to the decay width $H\to hh$ as a function
of the LO width for all points fulfilling condition (\ref{eq:cond3}) in accordance
with the experimental and theoretical constraints. It compares the
renormalization of $m_{12}^2$ through the process $H\to AA$ with the
one in the $\overline{\mbox{MS}}$ scheme with $\mu_R=2 m_h$. In both cases the
mixing angles are renormalized in the pOS scheme. Due to the large
involved couplings the corrections are found to be extremely large. In
the $\overline{\mbox{MS}}$ scheme the corrections are restricted to
values within about -300 and 150\% discarding small
LO widths. Corrections of this size can also be found in the
  process-dependent scheme, due to accidental cancellations among
  the various NLO terms. 
However, there are also scenarios yielding much larger relative
corrections with values beyond 600\% (not visible in the plot). In conclusion,  
the $\overline{\mbox{MS}}$ scheme is the preferable scheme for $m_{12}^2$
due to its better numerical stability that has been verified in the
investigations in the decoupling regime. Again, of course,
independent of the question of numerical stability, the overall large
corrections also in the process-independent schemes call for the
inclusion of higher order corrections that are beyond the scope of this paper.

\section{Conclusions \label{sec:concl}}
We investigated the renormalization of the mass parameter $m_{12}^2$,
which softly breaks the $\mathbb{Z}_2$ symmetry imposed on the 2HDM
Higgs potential. The impact of the renormalization through the 
$\overline{\mbox{MS}}$ scheme and through a process-dependent
definition via the decay $H \to AA$ was analyzed in the sample decay
$H\to hh$. 
While the process-dependent scheme cannot be tested in the decoupling 
regime and hence a statement on its numerical stability is prevented by
huge radiative corrections, our analysis still indicates an unfavourable
numerical behaviour of the process-dependent scheme when compared to the
$\overline{\mbox{MS}}$ scheme. The latter behaves better in the regime
where the loop corrections are dominated by strong coupling
contributions and the higher order corrections are hence
parametrically enhanced. Furthermore, it has proven good numerical properties
in the decoupling limit.
The Higgs decay into lighter Higgs pairs
also gave us the opportunity to reconfirm the good properties found
previously in the tadpole-pinched renormalization scheme for the
mixing angles $\alpha$ and $\beta$. Based on our findings we propose
for the renormalization of the 2HDM Higgs sector the application of
the tadpole-pinched scheme for the mixing angles $\alpha$ and $\beta$
and the $\overline{\mbox{MS}}$ scheme for $m_{12}^2$. These schemes lead
to manifestly gauge-independent counterterms, are process independent
and numerically stable. In scenarios featuring light CP-odd Higgs
bosons ($m_A \lsim 700$~GeV), the $p_\star^o$ scheme is less
preferable, due to the impact of the top resonance on $\delta \beta$ in this
scheme. 

\subsubsection*{Acknowledgments} 
The authors acknowledge financial support from the DAAD
project ``PPP Portugal 2015'' (ID: 57128671).  
Hanna Ziesche acknowledges financial support from the Graduiertenkolleg ``GRK
1694: Elementarteilchenphysik bei h\"ochster Energie und h\"ochster
Pr\"azision''. 
We are indebted to Marco Sampaio for kindly providing us with 2HDM data
sets. 

\vspace*{1cm}
\bibliographystyle{h-physrev}

\end{document}